\documentclass[notitlepage,11pt,eqsecnum,amsmath,preprintnumbers,superscriptaddress,nofootinbib,aps]
{revtex4-1}

\usepackage{hyperref}
\usepackage{multirow}
\usepackage{graphicx}
\usepackage{amsmath}
\usepackage{amssymb}
\usepackage{subfigure}
\usepackage{color,bm}

\long\def\del #1 \enddel { }

\usepackage{xcolor}
\usepackage{colortbl}

\definecolor{Gray}{gray}{0.85}
\definecolor{LightGray}{gray}{0.93}
\definecolor{LightGreen}{rgb}{0.88, 1, 0.88}
\definecolor{LightCyan}{rgb}{0.88,1,1}
\definecolor{LightRed}{rgb}{1, 0.85, 0.85}
\definecolor{LightYellow}{rgb}{1, 1, 0.85}
\definecolor{LightBlue}{rgb}{0.87, 0.94, 1}
\definecolor{white}{gray}{1}

\newcolumntype{G}{>{\columncolor{LightGray}}c}
\newcolumntype{L}{>{\columncolor{LightGray}}l}

\usepackage{graphicx}
\usepackage{amsmath}
\usepackage{amssymb}
\usepackage{subfigure}

\usepackage{epsfig}

\def\beq{\begin{equation}}
\def\eeq{\end{equation}}
\def\bea{\arraycolsep .1em \begin{eqnarray}}
\def\eea{\end{eqnarray}}
\def\Tr{{\rm Tr}}

\def\eq#1{(\ref{#1})}

\def\s0#1#2{\mbox{\small{$ \frac{#1}{#2} $}}}
\def\0#1#2{\frac{#1}{#2}}

\usepackage{array,mathtools,amssymb,booktabs}
\newcolumntype{C}{>{$}c<{$}}
\AtBeginDocument{
\heavyrulewidth=.16em
\lightrulewidth=.1em
\cmidrulewidth=.03em
\belowrulesep=.4ex
\belowbottomsep=0pt
\aboverulesep=.4ex
\abovetopsep=0pt
\cmidrulesep=\doublerulesep
\cmidrulekern=.5em
\defaultaddspace=.5em
}
\makeatletter
    \def\CT@@do@color{%
      \global\let\CT@do@color\relax
            \@tempdima\wd\z@
            \advance\@tempdima\@tempdimb
            \advance\@tempdima\@tempdimc
    \advance\@tempdimb\tabcolsep
    \advance\@tempdimc\tabcolsep
    \advance\@tempdima2\tabcolsep
            \kern-\@tempdimb
            \leaders\vrule
                    \hskip\@tempdima\@plus  1fill
            \kern-\@tempdimc
            \hskip-\wd\z@ \@plus -1fill }

   \makeatother
   
\begin{document}

\title{Fixed points and the spontaneous breaking of scale invariance}

\author{Daniel F.~Litim}
\affiliation{
\mbox{Department of Physics and Astronomy, University of Sussex, BN1 9QH, Brighton, UK}}
\author{Edouard Marchais}
\affiliation{
\mbox{Department of Physics and Astronomy, University of Sussex, BN1 9QH, Brighton, UK}}
\author{P\'{e}ter Mati}
\affiliation{
\mbox{Institute of Physics, Budapest University of Technology
  and Economics, H-1111 Budapest, Hungary}}
\affiliation{
\mbox{ELI-ALPS, ELI-Hu NKft, Dugonics t\'er 13, Szeged 6720, Hungary}}

\begin{abstract}
\centerline{\bf Abstract}
We investigate critical 
$N$-component scalar field theories and the spontaneous breaking of scale invariance  
in three dimensions using functional renormalisation. 
Global and local renormalisation group flows are solved analytically   in the infinite $N$ limit to establish the exact phase diagram of the theory including the Wilson-Fisher fixed point and 
a line of asymptotically safe UV fixed points.
We also study the  Bardeen-Moshe-Bander phenomenon of  spontaneously broken scale invariance and the stability of the vacuum for general regularisation.
Our findings  clarify a long-standing puzzle about the apparent unboundedness of the effective potential. Implications for other theories are indicated.
\end{abstract}
${}$\vskip-1.5cm

\pagestyle{plain} \setcounter{page}{1}

\maketitle

\tableofcontents

\section{\bf Introduction}

Fixed points of the renormalisation group play a fundamental role in quantum field theory and statistical physics \cite{Wilson:1971dc}. Low energy and infrared fixed points naturally control the long distance and low momentum behaviour of theories and are often associated with continuous phase transitions \cite {ZinnJustin:1996cy}. Ultraviolet (UV) fixed points  serve as a definition of quantum field theory, such as in asymptotic freedom \cite{Gross:1973id,Politzer:1973fx} or in asymptotic safety 
\cite{Weinberg:1980gg,Litim:2011cp,Litim:2014uca,Bond:2016dvk}. UV fixed points also ensure that the renormalisation group evolution of couplings remains finite even at highest energies. 
By their very definition,  fixed points imply that dimensionless couplings become independent of energy or length scale. Consequently, physical correlation functions  become scale-invariant, solely characterised by universal dimensionless numbers such as critical indices. 
An intriguing scenario arises in settings where quantum scale invariance at an interacting fixed point is broken spontaneously, leading 
to a theory whose mass scale is not determined by the fundamental parameters of the theory. It has also been speculated that this type of mechanism may offer a scenario for the origin of the Higgs as a ``light dilaton'' in certain extensions of the Standard Model \cite{Goldberger:2008zz,Bellazzini:2012vz}.

Spontaneously broken scale invariance at an interacting fixed point has first been observed by Bardeen, Moshe and Bander (BMB) \cite{Bardeen:1983rv,Bardeen:1983st} in strongly coupled 
$O(N)$ symmetric 
$(\phi^2)^3_{d=3}$ 
scalar field theories, and  in the limit of infinitely many fields $N$. At weak coupling, aspects of  symmetry breaking have been investigated for this model in \cite{Coleman:1974jh}, 
including $1/N$ corrections \cite{Townsend:1975kh} 
\cite{Townsend:1976sy}. The long distance behaviour and 
stability of the ground state have also been studied  \cite{Appelquist:1981sf,Appelquist:1982vd}.
Moreover, perturbative renormalisation group equations predict an interacting UV fixed point
\cite{Pisarski:1982vz}, and 
the role of composite operators $\sim \phi^2$ has also been investigated in view of 
vacuum stability  \cite{Gudmundsdottir:1984rr}. 
Using variational methods, however,
it emerged that 
perturbative  fixed points are unreliable  \cite{Bardeen:1983rv}. 
Rather, the phase diagram displays a line of fixed points whose tricritical endpoint features  the spontaneous breaking of scale invariance. Thereby a mass scale arises through dimensional transmutation from  the exactly marginal sextic scalar coupling \cite{Bardeen:1983st}. 
It has also been observed that $1/N$ corrections spoil the effect  \cite{David:1984we}. The model appears to have a viable ground state despite of  the BMB effective potential  being  unbounded from below \cite{David:1985zz}. 
Further aspects of the  ``BMB phenomenon'' were  analysed in \cite{Amit:1984ri,Ananos:1996kk} 
including extensions with global supersymmetry \cite{Bardeen:1984dx}, multicritical models \cite{Eyal:1996da} and critical scaling \cite{Comellas:1997tf}.
More recent investigations have dealt with the phase diagram of Wess-Zumino models \cite{Litim:2011bf,Heilmann:2012yf},  models with Chern-Simons gauge fields \cite{Bardeen:2014paa}, the phenomenon of ``walking'' 
\cite{Aoki:2014yra}, 
and the fate of light dilatons under $1/N$ corrections \cite{Omid:2016jve}.

In this paper, we are concerned with  interacting fixed points from the viewpoint of  the non-perturbative renormalisation group (RG)~\cite{Wilson:1973jj}.
The virtue of Wilson's setup is that it provides us with exact equations for running couplings and $N$-point functions following the successive integrating-out of momentum modes 
from a path integral representation of the theory. 
Various incarnations of ``functional renormalisation'' are available 
\cite{Polchinski:1983gv,Wetterich:1992yh,Morris:1993qb,Litim:2001up,Litim:2000ci} which, in combination with systematic approximations \cite{Golner:1985fg,Litim:1998nf,Berges:2000ew,Litim:2002ce,Pawlowski:2003hq,Blaizot:2005xy,Litim:2010tt},  give access to the relevant physics including at strong coupling. 
Recent applications of the methodology include   
models of particle physics \cite{Eichhorn:2015kea,Jakovac:2015iqa,Borchardt:2016xju}, purely fermionic models \cite{Jakovac:2013jua}, 
quantum gravity \cite{Litim:2003vp,Benedetti:2012dx,Demmel:2012ub,Dietz:2012ic,Falls:2013bv,Falls:2014tra},
 and models in fractal or higher dimensions \cite{Codello:2014yfa,Percacci:2014tfa,Mati:2014xma,Mati:2016wjn,Eichhorn:2016hdi,Kamikado:2016dvw}
The primary focus of this work will be on
$(\phi^2)^3_{d=3}$ 
scalar field theories in the limit of infinite $N$ \cite{PhysRev.86.821,Nicoll:1975yq}.
Our interest in this model is twofold:
Firstly,  the theory offers a rich spectrum of phenomena ranging from interacting ultraviolet fixed points with asymptotic safety to strongly interacting Wilson-Fisher fixed points, phase transitions and dimensional transmutation, and  interaction-induced spontaneous breaking of scale invariance \cite{Moshe:2003xn}. Secondly, the model can be solved analytically and offers important insights into the inner working of quantum  field theory. Additionally benefitting from recent findings in \cite{Litim:2016hlb,Juttner:2017cpr}, we thus provide a complete analytical solution of the theory covering all of its fixed points and for all RG scales $k$. Global and local renormalisation group flows are solved analytically  in terms of the microscopic parameters of the theory to establish the full phase diagram and universal scaling exponents.
We also investigate the  
spontaneously broken scale invariance and the generation of mass from the viewpoint of functional renormalisation for general regularisation, and explain how the mechanism is triggered through interaction-induced non-analyticities of the effective potential in the small field region. Our findings  resolve a long-standing puzzle by showing that the previously observed unboundedness of the effective potential is an artefact of a sharp momentum cutoff rather than a feature of the physical theory.  Whenever necessary, we also analyse the scheme (in-)dependence of results.

The paper is organised as follows. In Sec.~\ref{RG} we introduce the renormalisation group together with local and global flows and their exact solutions. In Sec.~\ref{FPU} we investigate  fixed points, scaling exponents, and the phase diagram. In Sec.~\ref{sBMB} we explain how scale invariance is broken spontaneously, also showing that the mechanism is universal. We summarize in Sec.~\ref{C}.

\section{\bf Renormalisation group}\label{RG}
In this section, we introduce our conventions and notation, the main equations for the theory at hand, as well as their local and global solutions. 

\subsection{Functional renormalisation}

Renormalisation group studies of $3d$ scalar field theories including at strong coupling   have a long history \cite{Wilson:1972qj}.  We are interested in the path integral quantisation of quantum field theory a la Wilson whereby  fluctuations are integrated out with the help of a  momentum scale parameter $k$ \cite{Wilson:1973jj,Polchinski:1983gv,Wetterich:1992yh,Ellwanger:1993mw,Morris:1993qb, Litim:2000ci}. Concretely, we consider Euclidean scalar field theories with partition function
\begin{equation}\label{Zk}
Z_k[J]=\int D\varphi\exp(-S[\varphi]-\Delta S_k[\varphi]-\varphi\cdot J)\,,
\end{equation}
where $S$ denotes the classical action and $J$ an external current. The Wilsonian cutoff term  
at momentum scale $k$ is given explicitly as
\beq
\Delta S_k[\varphi]=\frac12\int\,\frac{d^dq}{(2\pi)^d}\,\varphi(-q)\, R_k(q^2)\,\varphi(q)\,.
\eeq
The function $R_k$ obeys the limits $R_k(q^2\to 0)>0$ for $q^2/k^2\to 0$ and $R_k(q^2)\to 0$ for $k^2/q^2\to 0$ to guarantee that it acts as an infrared (IR) momentum cutoff \cite{Litim:2001up,Litim:2000ci,Litim:2001fd}. Moreover, the Wilsonian partition function \eq{Zk} 
falls back onto the full physical theory in the limit where the cutoff is removed $(k\to 0)$.
From \eq{Zk} the ``flowing'' effective action $\Gamma_k$ follows via a Legendre transformation $\Gamma_k[\phi]=\sup_J(-\ln Z_k[J]+\phi \cdot J) +\Delta S_k[\phi]$, where $\phi=\langle\varphi\rangle_J$ denotes the expectation value of the quantum field. The renormalisation group scale-dependence of $\Gamma_k$ is given by an exact functional identity \cite{Wetterich:1992yh} (see also \cite{Ellwanger:1993mw,Morris:1993qb}) as
\begin{equation}\label{FRG}
\partial_t\Gamma_k=\frac12\Tr\frac{1}{\Gamma_k^{(2)}+R_k}\partial_t R_k\,.
\end{equation}
It expresses the change with renormalisation group scale for the effective action $\Gamma_k$ in terms of an operator trace over the full propagator multiplied with the scale derivative of the cutoff itself.  We have also introduced the logarithmic flow parameter $t=\ln ({k}/{\Lambda})$, sometimes referred to as the  "RG-time". The presence of the momentum cutoff ensures that the flow is finite both in the UV and in the IR.  As such, the flow \eq{FRG} interpolates between  a microscopic  (classical) theory $(k\to\infty)$
and the full quantum effective action $\Gamma$ $(k\to 0)$,
\beq\label{k0}
\lim _{k\to 0}\Gamma_k=\Gamma\,.
\eeq
A few comments are in order. At weak coupling, iterative solutions of the flow \eq{FRG} generate the conventional perturbative loop expansion \cite{Litim:2001ky,Litim:2002xm}. It has also been established that the exact RG flow \eq{FRG} relates to the well-known Wilson-Polchinski flow~\cite{Polchinski:1983gv} by means of a Legendre transformation. 
The right-hand side (RHS) of the flow \eq{FRG} is  local in field and momentum space: for small momenta, the operator trace is suppressed owing to the momentum cutoff $R_k$ within the propagator. For large momenta, it is suppressed by the insertion $\partial_t R_k$, while for large fields the suppression arises through the field-dependent propagator itself.
This structure implies that the change of $\Gamma_k$ at momentum scale $k$ is local and governed by fluctuations with momenta of the order of $k$ \cite{Litim:2005us}. Locality is lost in the limit where the momentum cutoff $R_k(q^2)$ becomes a momentum-independent mass term whereby the flow \eq{FRG} reduces to the well-known Callan-Symanzik equation \cite{Litim:1998nf}.\footnote{In the Callan-Symanzik limit the RG flow is no longer UV finite and requires additional UV regularisation \cite{Litim:2006ag}.}
Optimised choices for the regulator term \cite{Litim:2000ci,Litim:2001up,Litim:2010tt} allow for analytic flows and an improved convergence of systematic approximations \cite{Litim:2005us}.

In the sequel, we are interested in $O(N)$ symmetric scalar field theory in $d$ euclidean dimensions to leading order in the derivative expansion, with the effective action
\begin{equation}\label{Gamma}
\Gamma_k=\int d^d x \left[ \frac{1}{2}(\partial\phi)^2 + U_{k} (\phi^a \phi_a) \right].\end{equation}
Here, $U_k$ denotes the dimensionful potential which depends on the $O(N)$ invariant terms of the theory, namely $\bar\rho\equiv\frac 12 \phi_a\phi ^a$.
The limit  $N\to \infty$ represents the universality class of the ideal Bose gas \cite{brezin1993large}. In this limit, the anomalous dimension of the field vanishes identically as the wavefunction does not get renormalized.  Moreover, provided the interaction Lagrangean solely depends on the invariant $\phi_a\phi ^a$ --- as is the case for the models investigated below --- the local potential approximation \eq{Gamma} becomes both exact  \cite{DAttanasio:1997yph}, and exactly soluble  
\cite{Tetradis:1995br,Litim:1995ex,DAttanasio:1997yph}. The reason for this is that derivative-type interactions are not switched on along the flow \eq{FRG} at infinite $N$. Their absence at any one scale $k_0$ (say, at the level of  the microscopic action) implies their absence at all scales $k$. In other words, the functional flow \eq{FRG} is closed for actions of the form \eq{Gamma},  which guarantees that the full effective action is given by \eq{Gamma} exactly,  for all $k$, and irrespective of the regularisation.\footnote{Conversely, in the presence of  derivative interactions,  the local potential approximation ceases to be exact and the effective action ceases to be exactly soluble, see~\cite{Morris:1997xj}
for an explicit example in $3d$ $O(N)$ symmetric models.} Below, we take  advantage of these facts for our analysis and conclusions. A discussion of $1/N$ corrections is  deferred until Sec.~\ref{1/N}.

By inserting the effective action from \eq{Gamma} into \eq{FRG}, and also introducing dimensionless variables
\begin{equation}
\begin{array}{rl}
u(\rho)&=U/k^d\\
\rho&=\frac12 \phi^2\,k^{2-d}
\end{array}
\end{equation}
 one finds the flow equation for the dimensionless potential
\begin{equation}\label{du}
\partial_t u = -du+(d-2)\rho u^\prime +(N-1)I[u']+I[u'+2\rho u'']
\end{equation}
in $d$ euclidean dimensions. The first two terms arise due to the canonical dimension of the potential and of the fields, whereas the third and fourth term arise due to the fluctuations of the $N-1$ Goldstone modes and the radial mode, respectively. The functions $I[x]$ encode the details of the Wilsonian momentum cutoff. They relate to the loop integral in \eq{FRG} and are given explicitly by 
\begin{equation}\label{Igen}
I[x]=k^{-d}\int d^dq\, \frac{\partial_t R_k(q^2)}{q^2+R_k+x\, k^2}\,.\end{equation}
In the present calculation we are mostly using the optimised regulator
\begin{equation}\label{opt}
R_{k}(q^2)=(k^2-q^2)\,\theta(k^2-q^2)\,,
\end{equation}
following \cite{Litim:2002cf,Litim:2001up,Litim:2000ci}. Then the integral \eq{Igen} in \eq{du} can be evaluated analytically, leading to 
\begin{equation}\label{I}
I[x]=\frac{A}{1+x}\,.
\end{equation}
The numerical factor $A=2/(d\,L_d)$ arises from the angular integration over loop momenta, and $L_d=(4\pi)^{d/2}\Gamma(d/2)$ denotes the $d$-dimensional loop factor \cite{Litim:2016hlb}. We have $A^{-1}=\s032 L_3$ with $L_3=4\pi^2$ in $d=3$ dimensions. At a fixed point solution, universal scaling exponents are independent of the numerical constant $A$.
In the limit of infinite $N$, the flow equation simplifies and we find
\begin{equation}\label{duprime}
\partial_t u'=-2u'+(d-2)\rho u''-\frac{u''}{(1+u')^2}\end{equation}
 for the flow of the first derivative of the potential.
Notice that we have  rescaled the field and the potential as $u\to u/A'$ and $\rho\to \rho/A'$ with $A'=N\cdot A$ to remove the loop factor and the matter multiplicity from the equations. The benefit of this  is that couplings are now measured in units of the appropriate loop factors 
following the rationale of naive dimensional analysis~\cite{Giudice:2003tu}.

\subsection{Local flows and exactly marginal coupling}
We now discuss the main features of the large-$N$ limit in the light of the underlying RG equations both locally, and globally. 
Expanding the flow equation \eq{duprime} in terms of the polynomial couplings at the potential minimum $u=\sum_n \lambda_n(\rho-\kappa)^n/n!$ where $\kappa(t)$ is the running minimum with $u'(\kappa)=0$. Using also $\lambda\equiv \lambda_2$, $\tau\equiv \lambda_3$, and writing $\beta_X\equiv dX/dt$ we find
\begin{eqnarray}
\beta_\kappa&=&1-\kappa\label{drho}\,,\\
\beta_\lambda&=&-\lambda(1-2\lambda)\,,\label{dlambda}\\
\beta_\tau&=&-6\lambda(\lambda^2-\tau)\,\label{dtau}
\end{eqnarray}
for the perturbatively relevant and marginal couplings. 
The local flows can be solved recursively for the fixed points of all higher couplings leading to explicit expressions for all $\lambda_n$ ($n>3$) in terms of the values  for $\kappa, \lambda$ and $\tau$ \cite{Litim:2016hlb}.
 We stress that the $\beta$-functions are non-perturbative, and exact. Two aspects determine the fixed point structure at infinite $N$. Firstly, the flow of the potential minimum  $\kappa$ factorizes from the remaining couplings. Hence, tuning $\kappa$ to its critical value 
\begin{equation}\label{kappaFP}
\kappa_*=\kappa_{\rm cr}=1
\end{equation} invariably leads to a fixed point solution. Its RG flow is UV attractive, implying that $\kappa$ is an IR relevant operator. Secondly, the quartic scalar coupling equally decouples from the system. Its RG flow \eq{dlambda} is independent of all other couplings, and displays two fixed points
\begin{equation}\label{lambdaFP}
\lambda_*=0\,,\quad{\rm or}\quad\lambda_*=\frac12\,.
\end{equation}
The first one is UV attractive, implying that it corresponds, together with \eq{kappaFP}, to a tricritical FP with two IR relevant couplings. The second one is IR attractive in $\lambda$, and corresponds to the Wilson-Fisher fixed point. Finally, the sextic coupling $\tau$ is classically marginal in three dimensions. Interestingly, for vanishing quartic coupling it becomes {\it exactly marginal}, see \eq{dtau}. Its fixed points are
\begin{equation}\label{tauFP}
\tau_*=\tau\,,\quad{\rm or}\quad\tau_*=\frac14\,.
\end{equation}
The exact non-renormalisation of the sextic interaction explains why its value, $\tau$, can be used to parametrize scaling solutions, provided that the quartic coupling vanishes, 
\beq
(\lambda_*,\tau_*)=(0,\tau)\,.
\eeq
For $\lambda_*\neq 0$, the subsystem of \eq{dlambda} and \eq{dtau} is closed and displays an exact non-perturbative and IR attractive fixed point 
\beq\label{eWF}
 (\lambda_*,\tau_*)=\left(\frac12,\frac14\right)\,.\eeq
The analytical solutions for the flow equations of the couplings read as follows
\begin{eqnarray}
\kappa(t)&=&1+c_\kappa\,e^{-t}\label{tkappa},\\
\lambda(t)&=&\frac{1}{2+c_\lambda\,e^t}\label{tlambda},\\
\tau(t)&=&\frac{2+c_\tau\,e^{3t}}{(2+c_\lambda\,e^t)^3}\label{ttau}.
\end{eqnarray}
The integration constants are determined by the initial values $\kappa_\Lambda, \lambda_\Lambda$ and $\tau_\Lambda$
\begin{eqnarray}
c_\kappa&=&-1+\kappa_\Lambda\label{ckappa},\\
c_\lambda&=&-2+1/\lambda_\Lambda\label{clambda},\\
c_\tau&=&-2+\tau_\Lambda/\lambda_\Lambda^3\label{ctau}.
\end{eqnarray}
As a final comment, it is interesting to have a look into the flow for the dimensionless mass term at fixed field $\rho=1$ representing the VEV,  $m^2\equiv u'(\rho=1)$, whose exact flow equation reads
\begin{eqnarray}
\beta_{m^2}&=&-2m^2 +\lambda-\frac{\lambda}{(1+m^2)^2}\,.\label{dm}
\end{eqnarray}
We notice that $m_*^2=0$ is always a fixed point, irrespective of the value of the quartic $\lambda$.\footnote{At the Wilson-Fisher fixed point \eq{eWF},
the apparent mass fixed points 
$m^2_{*,\pm}=\frac18\left(\pm\sqrt{17}-7\right)<0$ of \eq{dm}  
are spurious and do not extend to proper fixed points of $\Gamma_k$ for all fields.} Together with \eq{dlambda} we observe that both the mass and the quartic scalar self coupling remain unrenormalised, non-perturbatively, as soon as $(\kappa,m^2,\lambda)=(1,0,0)$.   This is in complete accord with earlier observations by Pisarski based on perturbation theory \cite{Pisarski:1982vz}.  We postpone a discussion of the phase diagram in Fig.~\ref{pBMB}  until Sec.~\ref{pd} below.

\subsection{Global flows and analytical solutions}
We now turn to a global characterisation of the RG flow diagram. The partial differential equation \eq{duprime} can be solved analytically by the methods of characteristics \cite{Litim:2016hlb}, see also \cite{Tetradis:1995br,Litim:1995ex}. For $u'\geq 0$, the explicit solution reads
\begin{equation}\label{+}
\frac{\rho-1}{\sqrt{u'}}-F(u')=G(u'e^{2t}),
\end{equation}
with 
\begin{equation}\label{F+}
F(u')=\frac{1}{2}\frac{\sqrt{u'}}{1+u'}+\frac{3}{2}\arctan\sqrt{u'}\,.
\end{equation}
The function $F(u')$ is bounded from above, $0\le F\le \s034 \pi$ for $u'\ge 0$, with $F(0)=0$.
The function $G(x)$ is fixed by the boundary condition $u'_\Lambda(\rho)$ at $k=\Lambda$ as 
\begin{equation}
\label{G}
G(x)=\frac{\rho_\Lambda(x)-1}{\sqrt{x}} -F(x)\,,
\end{equation}
with $\rho_\Lambda(u'_\Lambda(\rho))=\rho$. For $u'\leq 0$ the solution becomes 
\begin{equation}\label{-}
\frac{\rho-1}{\sqrt{-u'}}-\bar F(u')=\bar G(u'e^{2t}),
\end{equation}
where
\begin{equation}\label{F-}
\bar F(u')=-\frac{1}{2}\frac{\sqrt{-u'}}{1+u'}+\frac{3}{4}\ln\frac{1-\sqrt{-u'}}{1+\sqrt{-u'}}\,
\end{equation}
and suitably modified boundary condition, see (\ref{G}). The function $\bar F(u')$ is unbounded from below, $-\infty\le F\le 0$ for $-1\le u'\le 0$, with $\bar F(0)=0$.
 The branches  \eq{+} and \eq{-} and the functions (\ref{F+}) and (\ref{F-}) are related by analytical continuation in $u'$, using the relation
\[\frac{1}{i}\arctan{i x}=\frac{1}{2}\ln\left( \frac{1+ x}{1- x} \right).\]
together with the substitution $\sqrt{u'}\to i\sqrt{-u'}$
once $u'$ changes sign. 
\section{\bf Fixed points and universality}\label{FPU}

Fixed points are the scale-independent solutions of the RG flow. In this Section, we analyse the global fixed point solutions for the effective potential in three dimensions.

\begin{figure}[t]
\begin{center}
\includegraphics[scale=.35]{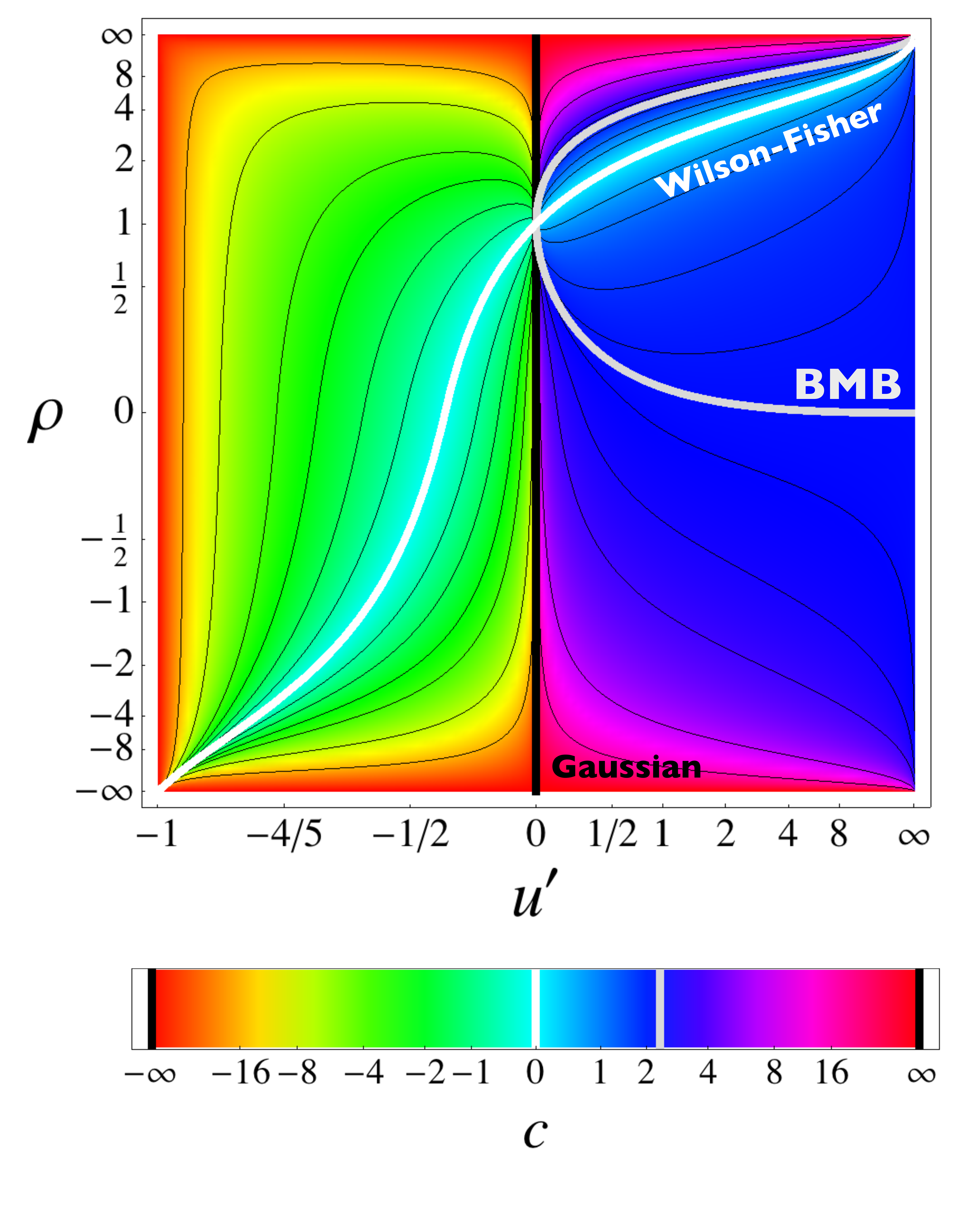}
\vskip-.75cm
\caption{Fixed point solutions $\rho(u')$ for all fields $\rho$ and all potentials $u'$, color-coded by the free parameter $c$. Thin lines correspond to fixed $c$ and are included to guide the eye, thick lines correspond to distinguished values for $c$ including the Wilson-Fisher fixed point (white line, $c=0)$, the BMB fixed point (gray, $|c|=c_P)$, see \eq{cP}, and the Gaussian fixed point (black, $|c|=\infty)$.  Axes are rescaled as $\rho\to\frac{\rho}{1+|\rho|}$ and $u'\to\frac{u'}{2+u'}$ for better display.}\label{ccode} 
\end{center}
\end{figure}

\subsection{Exact fixed points}
Fixed points of \eq{duprime} correspond to solutions \eq{+} and \eq{-} where the RHSs are independent of the RG scale $t$. By definition, they obey
\begin{equation}\label{duprime2}
0=-2u'+\rho u''-\frac{u''}{(1+u')^2}\,.
\end{equation}
From the fixed point equation \eq{duprime2} we deduce that all extrema or saddle points of the effective potential are all located at
\begin{equation}\label{min}
(\rho,u')=(1,0)\,,
\end{equation}
corresponding to the fixed point of the VEV \eq{kappaFP} detected within the local expansion. We also find that all continuously connected integral curves $\rho(u')$ can be classified in terms of a real parameter $c$. We chose the free real parameter $c$ to be given by
\begin{equation}\label{def+}
\left|\frac{\rho-1}{\sqrt{u'}}-F(u')\right|=c\,.
\end{equation}
 The regions with negative $u'$ are defined for $c\le 0$ via
\begin{equation}\label{def-}
\left|\frac{\rho-1}{\sqrt{-u'}}-\bar F(u')\right|=-c\,.
\end{equation}
Hence, all possible fixed point solutions are characterised by a real parameter $c$, which fixes
each and every point on the integral curve $\rho(u')$ via \eq{def+} or \eq{def-}. As we will see in more detail below, the absolute values in \eq{def+} and \eq{def-} arise to ensure that solutions are analytical around \eq{min}. For illustration, Fig.~\ref{ccode} shows a contour plot of all fixed point solutions \eq{def+} and \eq{def-}, colour-coded by the value for the free parameter $c$.
 
\subsection{Wilson-Fisher fixed point}
We now discuss the set of solutions in more detail. The limit $|c|\to\infty$ for either of the solutions \eq{def+} and \eq{def-}  implies that $u'$ vanishes identically, for all fields. This corresponds to the trivial Gaussian fixed point of the theory, given by the vertical middle line in Fig.~\ref{ccode}. 

\begin{figure*}[t]
\begin{center}
\includegraphics[scale=0.85]{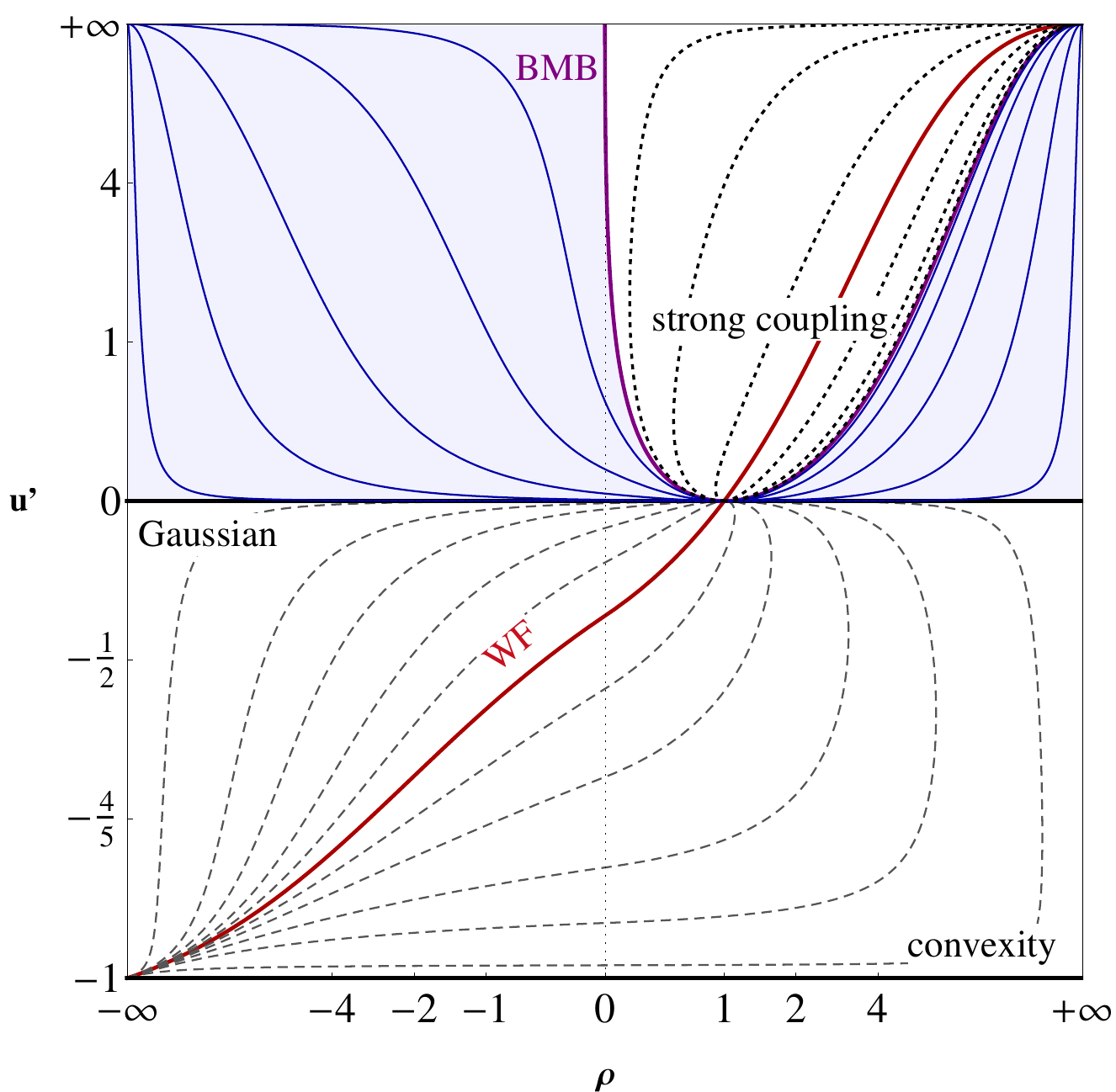}
\end{center}
\vskip-.5cm
\caption{Shown are all fixed point solutions $u'(\rho)$ of 3D $O(N)$ symmetric $(\phi^2)^3_{d=3}$ theories at infinite $N$. Full lines indicate physically viable solutions. Dotted and dashed lines indicate solutions with turning points in the regime with positive and negative $u'$, respectively. Special curves highlight the Wilson Fisher (WF), the Bardeen-Moshe-Bander (BMB), the Gaussian, and the convexity fixed point.
\label{fixed}}
\end{figure*}

For $c=0$, the fixed point solutions \eq{def+} and \eq{def-} are non-trivial, and, as we will argue, correspond to the seminal Wilson-Fisher fixed point. We have
\begin{equation}\label{WF}
\rho=1+H(u')
\end{equation}
where the function $H(x)$ is given by
\begin{eqnarray}\label{H}
H(x)\equiv\sqrt{x} \,F(x)=\sqrt{-x}\bar F(x)\,.
\end{eqnarray}
Notice that  both solutions \eq{def+} and \eq{def-} coincide due to \eq{H}. A few properties of $H$ are worth mentioning. It has an analytical expansion for small arguments,
\begin{eqnarray}\label{Hsmall}
H(x)&=&
2x-x^2+\frac45 x^3+{\cal O}\left({x^4}\right)\,.
\end{eqnarray}  
For asymptotically large arguments $x\gg 1$ we find
\begin{eqnarray}\label{Hlarge}
H(x)&=&
c_P\sqrt{x}-1+\frac1{5x^2}+{\cal O}\left({x^{-3}}\right)\,.
\end{eqnarray}  
Here, we have introduced the parameter
\begin{equation}\label{cP}
c_P=\frac{3}{4}\pi\,.
\end{equation}
For negative arguments $0<1+x\ll 1$, $H(x)$ displays a simple pole
\begin{eqnarray}\label{Hneg}
H(x)&=&
-\frac12\frac1{1+x} + \frac{1-3\ln 2}{2} +{\cal O}(1+x)\,.\quad
\end{eqnarray}  
Consequently, using \eq{Hsmall} in the vicinity of the minimum \eq{min} we find that the solution \eq{WF} becomes
\begin{equation}\label{WFsmall}
u'=\frac{1}{2}(\rho-1)+{\rm subleading}\,,
\end{equation}
where the subleading terms are higher order powers in $(\rho-1)$. Therefore the quartic scalar self-coupling reads $\lambda\equiv u''|_{\rho=1}=\s012$ at the potential minimum. Comparing with \eq{lambdaFP}
 we may conclude that the fixed point solution \eq{WF} must correspond to the Wilson-Fisher fixed point, and we denote \eq{WF}  as $\rho=\rho_{\rm WF}(u')$. Notice that $\rho_{\rm WF}(u')$ and the inverse function $u_{\rm WF}'(\rho)$ are monotonous function of $u'$ and $\rho$, respectively. Also, $u_{\rm WF}'(\rho)$ changes sign exactly once, as can also be seen from  Figs.~\ref{ccode},~\ref{fixed}. For asymptotically large fields, using \eq{Hlarge}, we have that
\begin{equation}
\sqrt{u'}=c_P \cdot\rho+{\rm subleading}
\end{equation}
showing that the asymptotic behaviour of $H$,\eq{cP}, determines the asymptotic growth of $u'$ with $\rho$.

\begin{figure*}[t]
\begin{center}
\hskip-1cm
\includegraphics[scale=.3]{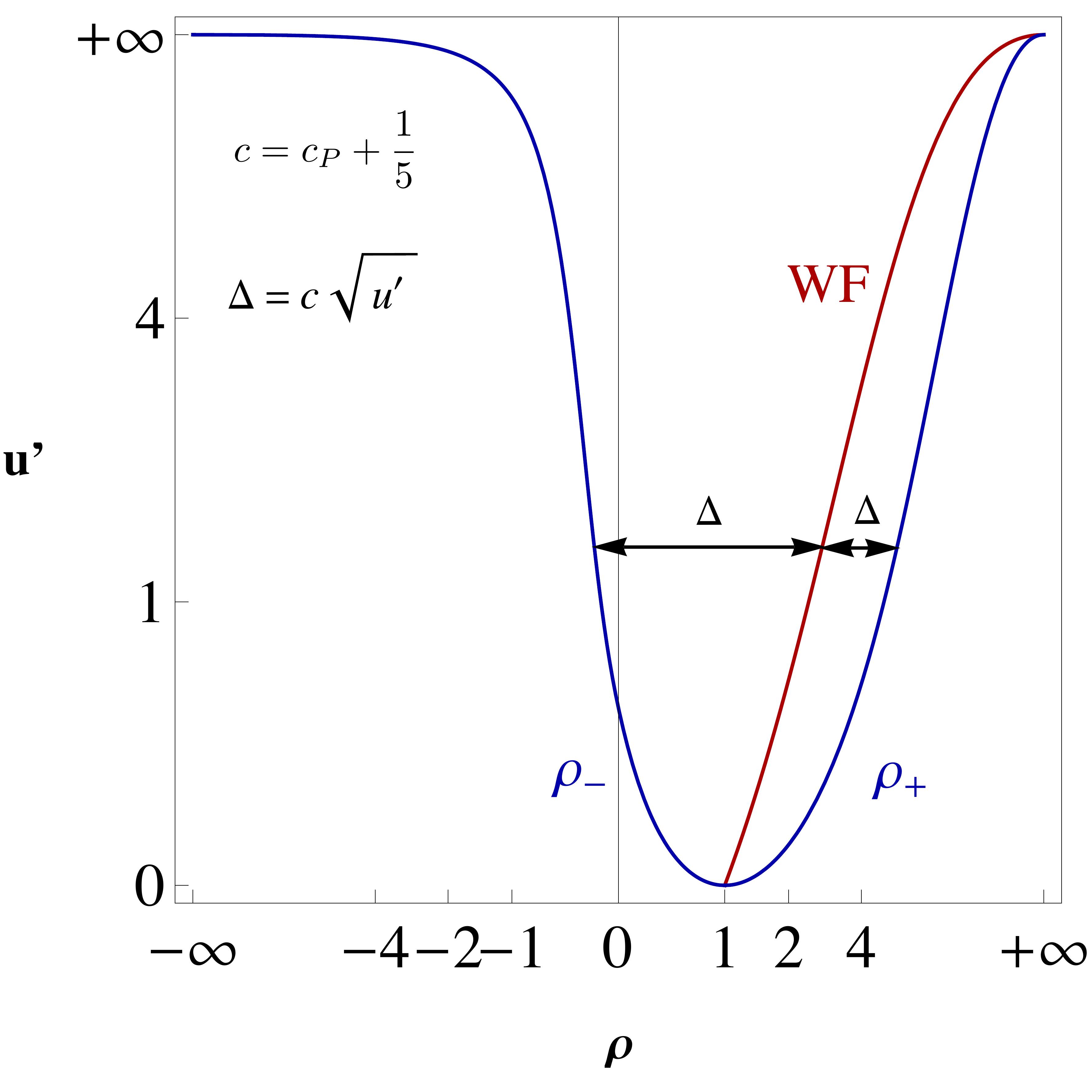}
\end{center}
\vskip-.5cm
\caption{Shown is a weak coupling fixed point with $c>c_P$ (thick blue line), together with the Wilson-Fisher fixed point where $c=0$. 
Notice that the two branches $\rho_\pm$ of the weak coupling fixed point follow from the Wilson Fisher fixed point by  a simple shift with $\Delta=c\sqrt{u'}$, see \eq{WF}, \eq{shift1}, where $c=c_{P}+\frac{1}{5}$.}
\label{branchweak}
\end{figure*}

\subsection{Tricritical fixed points}
Now we are in a position to discuss solutions with $c\neq 0$. From  Figs.~\ref{ccode} and~\ref{fixed}, we observe that integral curves may become multivalued unlike in the Wilson-Fisher case.\footnote{We refer to these as ``tricritical'' fixed points because they require the fine-tuning of two parameters (e.g.~the mass and the quartic). Conventional ``critical'' fixed points have a single relevant parameter (e.g.~the mass).} Starting with \eq{def+} for $c> 0$, the definition \eq{def+} can be resolved for $\rho=\rho(u')$ leading to two branches.  If
\begin{equation}\label{1a}
\rho\ge
1+H(u')
\end{equation}
then the branch is given by $\rho=\rho_+(u')$ with
\begin{equation}\label{1+}
\rho_+=1+H(u')+c \sqrt{u'}
\end{equation}
We recall that $H\ge 0$ in this regime. Conversely, if
\begin{equation}\label{1b}
\rho\le
1+H(u')
\end{equation}
then the branch reads $\rho=\rho_-(u')$ with
\begin{equation}\label{1-}
\rho_-=1+H(u')-c \sqrt{u'}\,.
\end{equation}
We conclude that the two branches for the field values $\rho=\rho_\pm$ follow from that of the Wilson-Fisher solution for positive $u'$ by a simple shift by $c\sqrt{u'}$ into either direction, \begin{equation}\label{shift1}
\rho=\rho_{\rm WF}(u')\pm c\sqrt{u'}\,.
\end{equation}

\begin{figure*}[t]
\begin{center}
\hskip-1cm
\includegraphics[scale=.3]{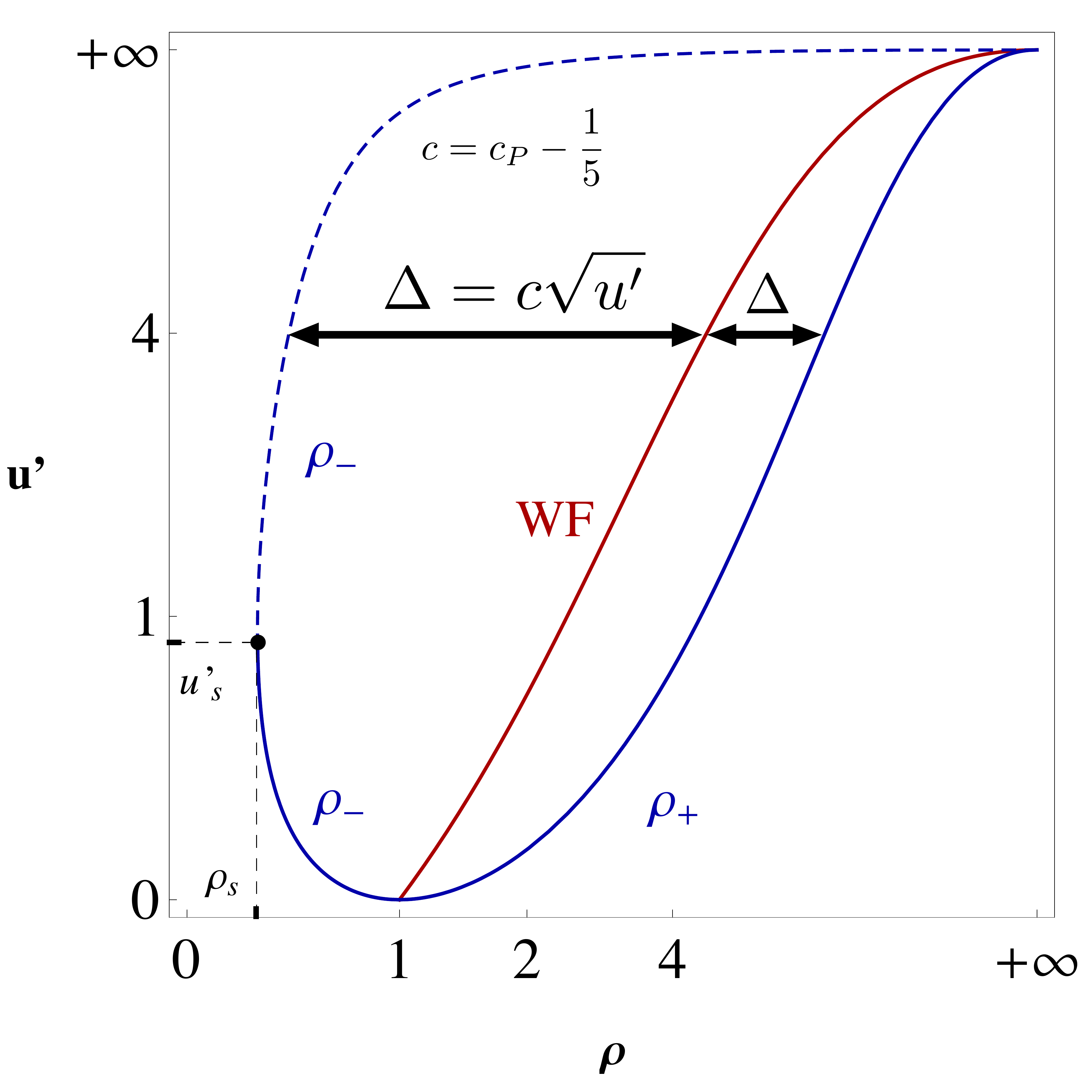}
\end{center}
\vskip-.5cm
\caption{Shown is a strong coupling fixed point for $0<c<c_P$ (full and dashed blue lines) together the Wilson-Fisher fixed point with $c_{\rm WF}=0$ (full red line).
Both branches $\rho_\pm$ of the strong coupling fixed point follow from the Wilson Fisher solution with positive $u'$ by a shift with $c\sqrt{u'}$, with $c=c_{P}-\frac{1}{5}$, see \eq{shift1}. Notice that  the strong coupling fixed point has become double-valued (indicated by the full vs the dashed line), 
characterised by a  turning point and a Landau-type singularity in its quartic self-interaction at $\rho=\rho_s$.}
\label{branch}
\end{figure*}

Both branches \eq{shift1} are continuously connected at \eq{min}. In fact, the values of the polynomial couplings in the vicinity of \eq{min} can be computed from either side, \eq{1+} and \eq{1-}, and must agree. We find
\begin{eqnarray}
u'&=&\s0{1}{c^2}(\rho-1)^2\,,\nonumber\\[1ex]
u''&=&\s0{2}{c^2}(\rho-1)\,,\label{tau}\\[1ex]
u'''&=&\s0{2}{c^2}\,,\nonumber
\end{eqnarray}
and similarly to higher order, modulo higher order corrections in $(\rho-1)$. The key point here is that the value of the sextic (and higher) interactions only depend on $c^2$, but not on the sign of $c$. This pattern guarantees that all higher couplings at \eq{min} are the same for \eq{1+} and \eq{1-}, ensuring that the solution is $C^\infty$ at \eq{min}. Furthermore, while $u'$ vanishes by definition at \eq{min}, we also find that the quartic interaction $\lambda$ is strictly vanishing at the minimum. This is the behaviour already found in  \eq{lambdaFP}. Hence, two couplings, the VEV and the quartic coupling, must be fine-tuned for this fixed point, which makes this a  tricritical fixed point. Comparing with \eq{tauFP}, we may link the free parameter $c$ with the exactly marginal sextic coupling $\tau$, finding
\begin{equation}\label{tauFull}
\tau=\frac2{c^2}\,.
\end{equation}
Notice that the sextic coupling $\tau$ is independent of the sign of $c$. 
Next we consider the large-field asymptotics of the solution. For either of the branches \eq{1+} and \eq{1-}, we find
\begin{equation}\label{asymp}
\sqrt{u'}=\frac{\rho}{c_P\pm c}\,.
\end{equation}
The significance of this result is that the large $u'$ asymptotics is controlled by $c_P\pm c$. For the branch \eq{1+}, the coefficient is always positive for all $c\ge 0$. For the branch \eq{1-} however, the coefficient changes sign at $c=c_P$. For $c>c_P$, the coefficient is negative and, consequently, \eq{asymp} describes the limit $\rho\to -\infty$. For $c<c_P$, the coefficient remains positive and \eq{asymp} becomes a large-field limit $\rho\to\infty$, similar to the result for \eq{1+} except for the proportionality factor. Hence the fixed point solution behaves qualitatively different depending on whether $c$ is larger or smaller than \eq{cP}. Examples for fixed point solutions with $c>c_P$ and $c<c_P$ are shown in Fig.~\ref{branchweak} and Fig.~\ref{branch}, respectively. 

\subsection{Landau-type singularity}
Around the VEV \eq{min} the branch \eq{1-} corresponds to fields $\rho<1$. To connect this branch with  fields $\rho\gg 1$ for  $c<c_P$, the integral curve  must display at least one turning point, see Fig.~\ref{branch}. At a turning point $d\rho(u')/du'=0$, and  the quartic coupling $u''(\rho_s)$  blows-up of at some point $\rho_s$ in field space. We obtain the location of turning points by rearranging \eq{duprime2} for $u''$. Assuming that $u_s'=u'(\rho_s)\neq 0$ at the turning point, we find

\begin{equation}\label{rhos}
\rho_s=\frac{1}{(1+u'_s)^2}\,.
\end{equation}
First we would like to know for which values of the parameter $c$ there is a turning point for the $u'\geq 0$ branch. It trivially follows from \eq{rhos} that $\rho_s \in (0,1]$. To figure out the exact values of the parameter $c=c_s$ for which a turning point at \eq{rhos} is achieved, we can insert \eq{rhos} into \eq{1-} to find
\begin{eqnarray}\label{cs}
c_s=\frac{1}{\sqrt{u'_s}}
\left(
u'_s\frac{u'_s+2}{(1+u'_s)^2}+H(u'_s)\right)\,.
\end{eqnarray}
This expression is monotonous in $u'_s$, with $u_s'$ within $[0,\infty)$, see \eq{rhos}. The expression is also bounded
\begin{equation}
\lim\limits_{u_s\rightarrow\infty} c_s=c_P\,,
\end{equation}   
thus confirming that for any value of $ c \in (0,c_P]$ the coupling $u''$ runs into a Landau-type pole at finite positive field. Landau-type singularities have previously been observed in the supersymmetric $O(N)$ theories \cite{Litim:2011bf,Heilmann:2012yf}, and for Wilson-Fisher fixed points in the complex field plane  \cite{Litim:2016hlb}.

\subsection{Fixed points with unbounded potentials}
We now repeat this analysis for the solutions \eq{def-} with $c< 0$ and $u'<0$. As we are going to show in detail below, these correspond to solutions with negative sextic coupling and to globally unstable effective potentials.

Solving \eq{def-} for $\rho=\rho(u')$ leads, again, to two branches. Notice that $H\le 0$ in this regime. If
$\rho\ge
1+H(u')$
the solution reads $\rho=\rho_+(u')$ with
\begin{equation}\label{2+}
\rho_+=1+H(u')+c \sqrt{-u'}\,.
\end{equation}
Conversely, if
$\rho\le
1+H(u')$
then the solution is given by $\rho=\rho_-(u')$ with
\begin{equation}\label{2-}
\rho_-=1+H(u')-c \sqrt{-u'}\,.
\end{equation}
The two branches for the field values $\rho=\rho_\pm$ follow from the Wilson-Fisher solution for negative $u'$  by a simple shift about $c\sqrt{-u'}$ into either direction, 
\begin{equation}\label{shift2}
\rho=\rho_{\rm WF}(u')\pm c\sqrt{-u'}\,.
\end{equation}
Both branches are continuously connected at \eq{min}. Using \eq{2+} and \eq{2-}, the polynomial couplings now read
\begin{eqnarray}
u'&=&-\s0{1}{c^2}(\rho-1)^2\,,\nonumber\\[1ex]
u''&=&-\s0{2}{c^2}(\rho-1)\,,\label{tau-}\\[1ex]
u'''&=&-\s0{2}{c^2}\,,\nonumber
\end{eqnarray}
and similarly to higher order. Comparing with \eq{tauFP} we may introduce the sextic coupling
\begin{equation}\label{tauFull-}
\tau=-\frac2{c^2}\,.
\end{equation}
There are two points worth noting. First, the left- and right-sided field derivatives at the potential minimum give exactly the same results, to all orders, establishing that the solutions \eq{def-} are indeed $C^\infty$. Second, we find that the sextic coupling has the opposite sign to \eq{tau}. Therefore the solutions \eq{def+} and \eq{def-} cover the entire range of exactly marginal sextic couplings.

\begin{table*}[t]
\centering
\begin{tabular}{cccc}
\toprule
\rowcolor{LightGreen}
&${}\quad$\bf  weak coupling${}\quad$  &${}\quad$\bf  critical coupling${}\quad$  &${}\quad$ \bf strong coupling${}\quad$\\ 
\rowcolor{LightGreen}
${}\quad$\bf range of field values${}\quad$&$(c>c_P)$&$\ \ (c=c_P)\ \ $&$\ \ (0\le c<c_P)\ \ $\\ \midrule
$1\le \rho\le \infty$&$\rho_+$&$\rho_+$&$\rho_+$\\
\rowcolor{LightGray}
$\rho\le 1$&$\rho_-$&&\\
$0\le \rho<1$&$$&$\rho_-$&$$\\
\rowcolor{LightGray}
$0<\rho_s\le \rho<1$&$$&$$&$\rho_-\ \  (u'<u'_s)$\\
$\rho_s\le \rho\le \infty$&$$&$$&$\rho_-\ \  (u'>u'_s)$\\ 
\midrule
\rowcolor{LightGreen}
${}$\quad\# of independent solutions${}\quad$&1&1&2\\
\bottomrule
\end{tabular}
\caption{Domains of validity for the fixed point solutions \eq{1+} and \eq{1-} with $u'\geq 0$. At weak and critical coupling, the junction of $\rho_+$ with $\rho_-$ provides a global fixed point solution $\rho(u')$. At strong coupling, two independent global solution exist for all fields $\rho\ge \rho_s$, see Figs.~\ref{branchweak} and \ref{branch} for weakly and strongly coupled examples, respectively.}\label{tab:rel}
\end{table*}

For the same reason it is  not possible to analytically connect branches with positive $u'$ to those with negative $u'$ for any finite, non-zero $c$. Although the first two couplings would trivially agree, from the sextic coupling onwards we would observe discontinuities. In other words, for $c\neq 0$, a fixed point solution which has a specific sign for $u'$ at some field value cannot change its sign along the entire integral curve.

For negative $u'$, the fixed point solution may display a turning point \eq{rhos}. It is located at $\rho_s\in (1,\infty)$ for $u'_s\in (0,-1)$. In our conventions, a turning point at $u'_s<0$ and with $\rho_s$ given by \eq{rhos} corresponds to the parameter $c=c_s$, with
\begin{eqnarray}\label{csm}
c_s=-\frac{1}{\sqrt{-u'_s}}
\left|
u'_s\frac{u'_s+2}{(1+u'_s)^2}+H(u'_s)\right|
\,.
\end{eqnarray}
We notice that $c_s=0$ for $u'_s\to 0$. In the opposite limit where $u'_s\to-1$, \eq{csm}  states that $c_s\to-\infty$, and $\rho_s\to\infty$. This exhausts all possible values for $c$ in \eq{def-}. Hence, for each and every finite coupling $c<0$, the solution \eq{def-} displays a turning point at positive fields.   In particular, solutions cannot be extended to arbitrarily large fields $\rho>0$. 
For this reason, we conclude that none of these solutions are physically viable, globally.
This is in marked contrast to solutions with positive $u'$. The only solution which would extend over all fields is the limit $1/c\to 0^-$. This limit is responsible for the approach towards convexity of the effective potential in a phase with spontaneous symmetry breaking. Some fixed point solutions in either regime are displayed in Fig.~\ref{fixed}.

\subsection{Line of ultraviolet fixed points}
We may summarize our findings as follows. The tricritical fixed point solutions of $3d$ scalar field theories in the infinite $N$ limit are characterized by a vanishing mass term $m^2=0$, a vanishing quartic coupling $\lambda=0$ and an exactly marginal sextic coupling $\tau$ in the vicinity of the minimum \eq{min}. 
For positive sextic coupling $\tau=\frac2{c^2}$, we find several qualitatively different regimes: In the {\it weak coupling regime} where
\begin{equation}\label{weak}
c_P< c\quad{\rm and}\quad\tau< \frac2{c_P^2}
\end{equation}
the fixed point solution is single-valued and well-defined for all fields. The scaling solution always displays a local extremum $u'=0$. This regime  includes the free Gaussian fixed point in the limit
\begin{equation}\label{cGauss}
 |c|\to \infty\quad{\rm and}\quad\tau\to 0\,,
 \end{equation}
consistent with \eq{weak}. The latter entails that all higher order couplings vanish identically at the Gaussian fixed point.  Inasmuch as the Gaussian corresponds to a free UV fixed point,
the fixed points with \eq{weak} correspond to a line of interacting, asymptotically safe, ultraviolet fixed points.
In the {\it strong coupling regime},
\begin{equation}\label{strong}
c< c_P\quad{\rm and}\quad\tau> \frac2{c_P^2}
\end{equation}
the theory develops a Landau-type singularity in $du'/d\rho|_{\rho_s}$ at some finite field value in the physical regime. Furthermore, for fields with $0<\rho<\rho_s$ the effective potential is not defined. Strictly speaking the fixed point solution does not exists for all fields. It is conceivable that the theory develops bound states, or ground states different from constant fields, as tacitly assumed here. Interestingly, at strong coupling the theory also displays two independent scaling solutions corresponding to the same parameter value $c<c_P$. Only one of them displays a local minimum, and only for that one the parameter $c$ is related to the exactly marginal sextic coupling $\tau$. This type of degeneracy of solutions at strong coupling has been noticed previously in the context of the supersymmetric $O(N)$ theory \cite{Litim:2011bf,Heilmann:2012yf}. 
 An interesting special case is provided by the borderline between strong and weak coupling, which we refer to as the {\it critical coupling}
\begin{equation}\label{BMB}
c=c_P\quad{\rm and}\quad\tau= \tau_P\equiv \frac2{c_P^2}\,.
\end{equation}
Here, the effective potential becomes non-analytic at $\rho=0$ and displays the Bardeen-Moshe-Bander phenomenon. We return to this in more detail in Sec.~\ref{sBMB}. 
It is worth noticing  that the appearance of the critical coupling strength \eq{BMB} is not directly  visible from within the local expansion. In fact, the local expansion would find a fixed point even in the strong coupling regime, but more work is required to notice that these reach a singularity in the quartic coupling for sufficiently small fields. This is reminiscent of earlier fixed point  studies within perturbation theory \cite{Pisarski:1982vz}.

Finally, we observe that the  Wilson-Fisher fixed point of $3d$ scalar field theories in the infinite $N$ limit corresponds to the parameter value
\begin{equation}\label{cWF}
 c=0\,.
\end{equation}
It corresponds to an isolated fixed point which is not continuously connected to the tricritical line of UV fixed points.  In view of \eq{strong}, the theory with \eq{cWF} would corresponds to the strong-coupling regime where the sextic coupling takes a value of order unity. In addition, the Wilson-Fisher fixed point differs from the line of tricritical fixed points in that the quartic self-coupling  is non-vanishing, equally taking a value of order unity \eq{lambdaFP}.

No viable fixed point solutions are found for the branch with negative sextic coupling where $u'$ stays negative for all fields. The corresponding potentials are unbounded, and do not exist beyond a critical field value, irrespective of the coupling strength.

\subsection{Scaling exponents from local flows}

Scaling exponents describe how the physical observables behave cloe to a continuous phase transition. Within functional renormalisation, they have been computed up to fourth order within the derivative expansion using local and global flows including error estimates~\cite{Litim:2010tt}. While exponents depend on the RG scheme within fixed approximations \cite{Litim:2007jb}, they come out RG scheme independent at infinite $N$, $e.g.$~\cite{DAttanasio:1997yph,Morris:1997xj,Comellas:1997tf,Litim:2010tt,Litim:2001dt}.

Universal scaling exponents can be extracted from the RG equations in many ways. If we use a the polynomial approximation $u=\sum_n \lambda_n(\rho-\kappa)^n/n!$ (which is definitely well justified close to the minimum) we can get the critical exponents from the linearized beta-functions evaluated at the fixed points. In fact, we need to solve the eigenvalue equation for the stability matrix
\begin{equation}
Bv^I=-\theta^Iv^I\,,
\end{equation}
where $B_{mn}\equiv\partial \beta_m/\partial \lambda_n|_{\lambda=\lambda_*}$ denotes the stability matrix, $v^I$ its eigenvectors, and $\theta^I$ the corresponding eigenvalues. 
In critical phenomena, the exponent $\nu$ denotes the divergence of the correlation length $\bar m\propto|T-T_c|^{\nu}$ with temperature $T$. $T_c$ denotes the ``critical temperature'' whose role is taken here by the critical VEV. Then $\nu$ is given by $\nu=-1/\theta$, where $\theta$ denotes the most negative eigenvalue amongst the eigenvalues of $B$.

The relevant and marginal scaling exponents are encoded in the local flows \eq{drho}, \eq{dlambda} and \eq{dtau}. 
Notice that the system \eq{drho} --\eq{dtau} does not contain  the mass term explicitly. At the Gaussian fixed point, it would thus display the eigenvalues $\{-1,-1,0\}$ corresponding to minus the canonical mass dimension of the couplings. In order to detect Gaussian exponents we replace the running of the VEV $\kappa$ \eq{drho} by the running of the mass term $m^2$ at fixed field $\rho=1$ whose exact flow equation has been given in \eq{dm}.
The flows \eq{dlambda}, \eq{dtau} and \eq{dm} now display the canonical eigenvalues $\{-2,-1,0\}$ as expected for the Gaussian fixed point. The corresponding index $\nu$ takes its mean field value $\nu=\s012$. This eigenvalue pattern with $\theta_n=n-2$ for $n\ge 0$ persists to higher order in a polynomial expansion.

Next we consider the tricritical fixed points, with  $(\kappa=1,\lambda=0,\tau=2/c^2)$ for finite $c\neq 0$.The fixed point solution where $c$ is an arbitrary parameter has been defined above. Calculation of the eigenvalues from \eq{dlambda}, \eq{dtau} and \eq{dm} gives us canonical eigenvalues
\begin{equation}\label{thetaTri_local}
\theta^I\in\{-2,-1,0,\cdots\}\,.
\end{equation}
This pattern persists to higher order in a polynomial expansion. Two aspects are worth noticing. Firstly, the theory displays classical (Gaussian) exponents, yet the fixed point is interacting and non-Gaussian. Secondly, the local derivation of exponents is unable to differentiate  the regimes of strong and  weak coupling. Locally, these scaling exponents are found for all sextic coupling. From the global analysis of the flow, however, we may conclude that the result is only valid in the region $c>c_P$ (i.e. for weak coupling). 
Finally, turning to the Wilson-Fisher fixed point,
we find the following eigenvalues in a local polynomial expansion,
\begin{equation}\label{thetaWF_local}
\theta^I\in\{-1,1,3,\cdots\}\,.
\end{equation}
The sole negative eigenvalue implies $\nu=1$.

 \subsection{Scaling exponents from global flows}
In the infinite-$N$ limit of the $O(N)$ symmetric model, the critical exponents and the global linear eigenperturbations can be calculated exactly without resorting to a polynomial approximation \cite{Comellas:1997tf,Litim:2011bf,Heilmann:2012yf}. To that end, we consider small perturbations around the full solution of the flow \eq{FRG} in the vicinity of the fixed point
\begin{equation}
u'(t,\rho)=u'_{*}(\rho)+\delta u'(t,\rho)\,.
\end{equation}
After  linearization of the flow  for $\delta u$ one finds  
\begin{equation}\label{flu}
\partial_{t}\,\delta u'=
2\frac{u_*'}{u_*''}\left(\partial_\rho-\frac{(\sqrt{u_*'}\cdot u_*'')'}{\sqrt{u_*'}\cdot u_*''}\right)
\delta u'\,.
\end{equation}
Eigenperturbations with eigenvalue $\theta$ obey
\begin{equation}\label{dtheta}
\partial_t \,\delta u'=\theta\, \delta u'\,,
\end{equation}
and \eq{flu} with \eq{dtheta} can be solved by separation of variables. We write $\delta u'(\rho,t)=T(t)R(\rho)$ to find
\begin{eqnarray}
(\ln T)'&=&\theta\,,\\
(\ln R)'&=&\s012 \theta \,(\ln u'_*)'+\s012(\ln u'_* (u''_*)^2)'\,.
\end{eqnarray}
These are integrated  to give
\begin{eqnarray}
T(t)&\propto&e^{\theta t},\\
R(\rho)&\propto& u_*'' (u_*')^{\frac{1}{2}(1+\theta)}.
\end{eqnarray}
We conclude that eigenperturbations are given by
\begin{equation}\label{epert}
\delta u' =C \,e^{\theta t}(u_*')^{\frac{1}{2}(1+\theta)}u_*''
\end{equation}
to linear order, where $C$ is an unspecified normalisation constant. 

We now discuss the quantisation of eigenvalues $\theta$. The allowed set of values for the exponent $\theta$ is determined by imposing analyticity conditions for the eigenperturbations. We begin with the scaling exponents for the Wilson-Fisher fixed point. From \eq{WFsmall} we infer that $u'_*$ is linear in $(\rho-1)$ around the minimum and $u''_*$ a constant, and hence
\begin{equation}\label{EP_WFsmall}
\delta u' \propto   e^{\theta t}(\rho-1)^{\frac{1}{2}(1+\theta)}\,.
\end{equation}
For asymptotically large fields, eigenperturbations scale as
\begin{equation}\label{EP_large}
\delta u' \propto   e^{\theta t}\rho^{(2+\theta)}\,.
\end{equation}
Requiring that  eigenperturbations are analytical functions of the field $\rho$ in the vicinity of the potential minimum imposes that the power $\frac{1}{2}(1+\theta)$ takes positive integer values, thus
\begin{equation}
\theta=-1,1,3,5,7 \cdots
\end{equation}
in agreement with \eq{thetaWF_local}. Notice that both the small and the large field asymptotic behaviour are integer powers of the field. 

For the tricritical fixed points, we have established in \eq{tau} that $u'_*$ is quadratic around the minimum, while $u''_*$ is linear. 
One may therefore rewrite the RHS of \eq{epert} in terms of $\sqrt{u_*'}$ which is linear in the field around the minimum and whose first derivative becomes a constant.
This leads to 
\begin{equation}\label{EVsqrt}
\delta u'=C \,e^{\theta t}(\sqrt{u_*'})^{2+\theta}\left(\sqrt{u_*'}\right)'
\end{equation}
Comparing \eq{EVsqrt} with \eq{epert}, we conclude that imposing analyticity conditions now requires $\theta+2$ to be a positive integer,
\begin{equation}\label{tri}
\theta =-2,-1,0,1,2,3,\cdots\,,
\end{equation}
in agreement with \eq{thetaTri_local}. Close to the minimum, eigenperturbations scale as
\begin{equation}\label{EVtri}
\delta u' \propto e^{\theta t}(\rho-1)^{\theta+2}\,.
\end{equation}
while for asymptotically large fields the behaviour is given by \eq{EP_large}. As an aside, the result \eq{EVtri} shows that the theory displays an interacting UV fixed point with exact Gaussian exponents. This is an exact example for the general behaviour of scaling exponents conjectured in \cite{Falls:2013bv}, according to which operators with increasing mass dimension become increasingly irrelevant at interacting fixed points. Near-Gaussian exponents at UV fixed points have recently been observed in certain models of 4d quantum gravity \cite{Falls:2014tra}.

As a final comment, we note that the eigenperturbations for the potential $\delta u$ follow from our results \eq{epert} and \eq{EVsqrt} by direct integration with respect to $\rho$. We find
\begin{equation}\label{EVpot}
\delta u=C' \,e^{\theta t}(\sqrt{u_*'})^{3+\theta}
\end{equation}
for the cases where $c\neq 0$, and
\begin{equation}\label{EVpot_WF}
\delta u=C' \,e^{\theta t}(u_*')^{\frac{1}{2}(3+\theta)}
\end{equation}
for $c=0$. In either case, we note the appearance of a new eigenvalue $\theta=-3$ corresponding to  shifts in the zero point energy. However, in the absence of (quantum) gravity, this eigenvalue is irrelevant despite its negative value because the vacuum energy does not influence the location of the fixed point.

\begin{figure}[t]
\begin{center}
\hskip-1.5cm
\includegraphics[scale=.35]{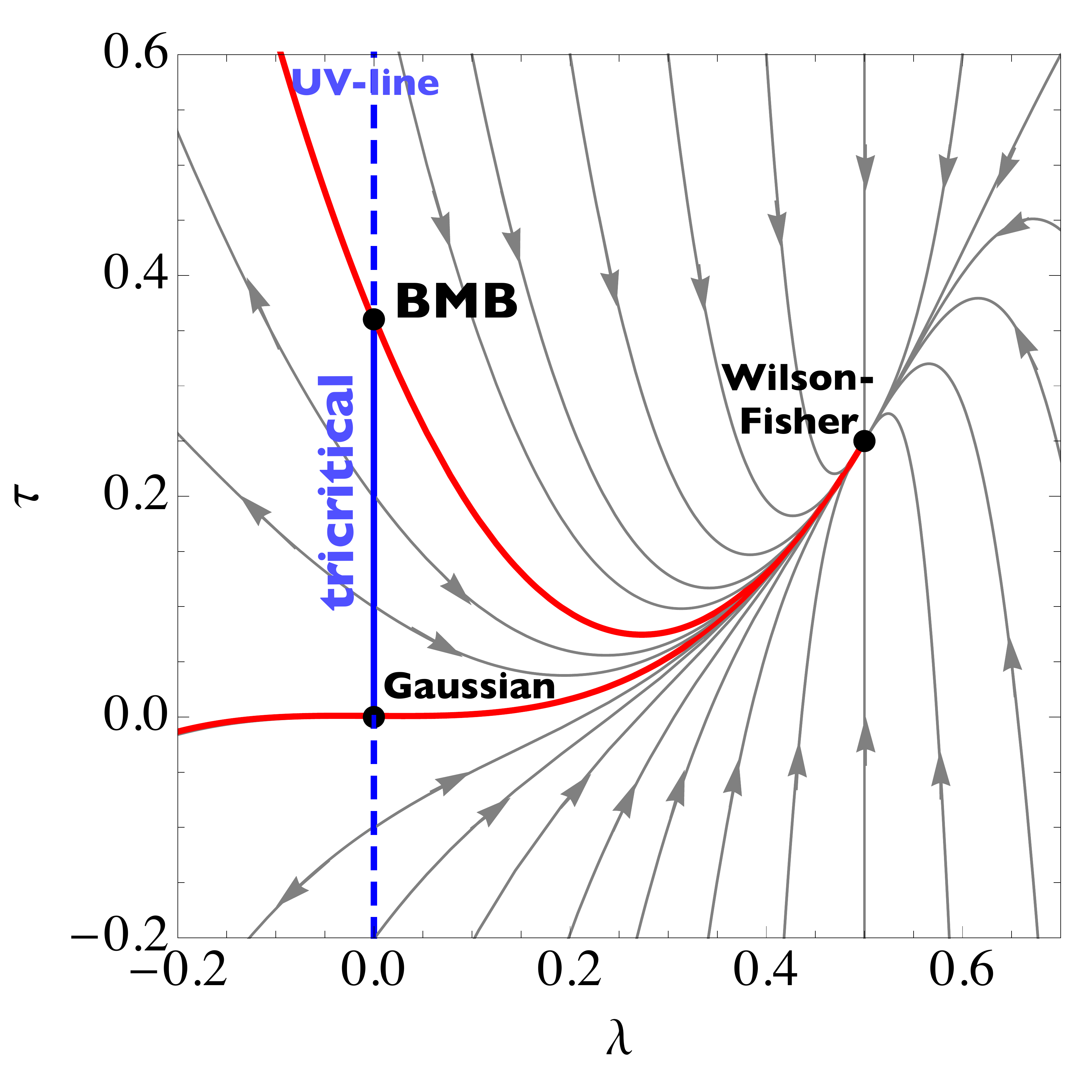}
\caption{Shown is the exact phase diagram of $(\phi^2)^3_{d=3}$-theory at infinite $N$ within the plane of the quartic and sextic coupling $(\lambda,\tau)$. Arrows indicate the flow towards the IR. The tricritical line of fixed points (full blue line), as well as the Gaussian, the Wilson-Fisher (WF), and the Bardeen-Moshe-Bander (BMB) fixed points are indicated (black dots). Trajectories are emmenating from the UV-line $(\lambda=0)$. The separatrices connecting the BMB and the Gaussian FPs with the WF FP are shown in red. The WF fixed point takes the role of an IR attractor.  The flow of the vacuum expectation value \eq{drho} (not shown) adds an IR  unstable direction to the phase diagram; all higher order polynomial interactions are IR stable. The full (dashed) blue line segment relates to models with (without) well-defined fixed point potential for all (dimensionless) fields (see main text).}\label{pBMB} 
\end{center}
\end{figure}
\subsection{Phase diagram and asymptotic safety}\label{pd}On the level of the global phase diagram  for the $(\phi^2)^3_{d=3}$ theory at infinite $N$, our results are summarised in Fig.~\ref{pBMB}.
It shows the RG flows in the plane of the quartic and sextic couplings $(\lambda,\tau)$ defined at the local minimum of the effective potential, and on the hypersurface with $\kappa=\kappa_*$. The tricritical line of asymptotically safe UV fixed points (full blue line), as well as the Gaussian (G) and the Wilson-Fisher (WF) fixed points are indicated. The endpoint of the tricritical line corresponds to the BMB fixed point, whose study is detailed Sect.~\ref{sBMB}. The dashed blue line indicates strongly-coupled fixed points without a global potential where either $\tau>\tau_P$ or $\tau<0$. Arrows indicate the flow towards the IR, and the separatrices connecting the BMB and the Gaussian fixed point with the WF one are highlighted as well (thick red lines). In the  $(\lambda,\tau)$-plane, the WF fixed point takes the role of an IR attractor. Trajectories are emmenating from the UV-line $(\lambda=0)$, along which the sextic coupling is exactly marginal. The flow of the vacuum expectation value, \eq{drho}, has decoupled from all other couplings. It adds an IR  unstable direction to the entire phase diagram including the Wilson-Fisher fixed point.  Away from the critical hypersurface of strictly massless theories $(\kappa\neq\kappa^*)$ the flow is driven via first or second order phase transitions towards the symmetric phase $(\bar \kappa=0)$ and the symmetry broken phase $(\bar \kappa>0)$ in the IR.
  
  The line of tricritical fixed points corresponds to exact interacting UV fixed points. All trajectories emanating from it can be viewed as well-defined UV-complete microscopic theories. In this light,  for all tricritical fixed points with $0\le\tau\le \tau_P$ the theory is asymptotically safe in the spirit of Weinberg's asymptotic safety conjecture \cite{Weinberg:1980gg}. Exact asymptotic safety arises as a consequence of an exactly marginal  coupling $\tau$, allowing a continuous interpolation between asymptotic freedom  at the Gaussian UV fixed point for  $\tau\to 0$ to asymptotic safety along the line of interacting UV fixed points for $0<\tau\le \tau_P$.\footnote{Exact asymptotic safety and vacuum stability has also been established in four dimensional theories (without gravity), see  \cite{Litim:2014uca,Litim:2015iea,Bond:2016dvk}, and  \cite{Bond:2017wut} for recent applications in particle physics.}

\subsection{Discussion}

We have provided a complete analysis of fixed points for the $O(N)$-symmetric scalar theories in the large-$N$ limit. A central role is played by the scalar sextic coupling, which becomes exactly marginal provided quartic interactions are absent. In this regime the theory also displays two relevant couplings given by the VEV and the quartic interactions. For weak positive sextic coupling, the theory displays unique and well-behaved asymptotically safe UV fixed points and classical scaling exponents. For strong positive sextic coupling, the fixed point bifurcates into two solutions both of which exist for large, but no longer for small fields. For both of these, the unavailability of an effective potential for small fields is signalled by an integrable  Landau-type singularity in the quartic interactions.  In this regime the ground state is no longer well-described by constant fields. Interestingly though, both at weak  and strong sextic coupling, the theory displays classical scaling with $\nu=\s012$
despite of the fact that the theory is interacting. 

For negative sextic coupling, whether large or small, no global fixed point is found. Although a local solution exists for small fields, the would-be fixed points terminate through a Landau-type singularity and the effective potential ceases to exist. Furthermore, the effective potential is unbounded as $u'<0$ in its domain of validity. We conclude that physically admissible fixed points do not exist for negative sextic coupling. 

Non-classical scaling with $\nu=1$ is found at the unique Wilson-Fisher fixed point, where the quartic coupling is non-vanishing.\footnote{Non-classical scaling with $\nu=\s013$ arises at the endpoint of the line of tricritical fixed points
\cite{Bardeen:1983rv}.}    While the classical and non-classical scaling exponents arise correctly from both the local and the global RG flows, the qualitative difference between weak and strong coupling including the existence of an endpoint of the line of critical points and the non-availability of an effective potential for small fields, only became visible through the study of the global RG flows. The local RG flows, based on a polynomial approximation of the effective action, correctly show critical and tricritical fixed points, but they do not offer direct indications for the critical endpoint. This pattern is reminiscent of perturbative studies at large-$N$, which failed to identify the endpoint. We conclude that in order to uncover the onset of strong coupling behaviour signalled by the critical end point, non-perturbative methods beyond polynomial expansions or conventional perturbation theory must be adopted.

\section{\bf Spontaneous breaking of scale invariance}\label{sBMB}

In this section, we turn to the critical end point and the spontaneous breaking of scale invariance, and establish how it arises universally using the method of functional renormalisation.

\subsection{Fixed point and the origin of mass} \label{SBSI}
By their very definition, fixed points of the renormalisation group imply that dimensionless couplings remain unchanged under a change of RG momentum scale, which entails scale invariance. Interestingly, in \cite{Bardeen:1983rv} it was noticed  that scale invariance may nevertheless be broken spontaneously at a fixed point. From a renormalisation group perspective, this can be understood as follows. Consider a fixed point solution $u_*(\rho)$ for the effective potential. The physical mass  $m$ of the scalar field in the symmetric phase at the fixed point is then given by
\begin{equation}\label{mPhys}
m^2=u_*'(0)\,k^2\,.
\end{equation}
Notice that $m^2$ is non-zero for $u_*'(0)\neq 0$ and $k\neq 0$. This is in accord with scale invariance because the mass parameter scales proportionally to the RG scale, $m\sim k$. Moreover, as long as $u_*'(\rho=0)$ at vanishing field remains finite, the physical mass vanishes in the physical limit \eq{k0} where all fluctuations are integrated out,
\begin{equation}
m^2\to 0\,,
\end{equation} 
as one would expect for a scale-invariant theory. This is genuinely true for any quantum field theory irrespective of the sign of $u'_*(0)$. In our case, this applies for all weakly coupled fixed points studied in Sec.~\ref{FPU}, where $c> c_P$. We stress, however, that this conclusion centrally relies on $u'_*(0)$ being finite.  It may be upset provided that $u_*'(0)$ diverges, 
\begin{equation}\label{uprime0}
u'_*(0)\to\infty\,,
\end{equation}
in which  case the physical mass \eq{mPhys} may take any finite value, 
\begin{equation}
m^2\to {\rm finite}\,,
\end{equation} 
and remains undetermined otherwise in the limit $k\to 0$. We observe that the mass has become a free parameter which is not determined by the fundamental parameters of the theory \cite{Bardeen:1983rv,David:1984we}.
We conclude that theories with a diverging $u_*'(0)$ at a fixed point may lead to the spontaneous breaking of scale invariance due to the dynamical generation of a mass scale.

\subsection{Dimensional transmutation and non-analyticity}
We now turn to an analysis of the breaking of scale invariance for $O(N)$ theories at infinite $N$. Firstly,  within the set of all scaling solutions  parametrised by $c$ we notice that a diverging $u_*'(0)$ can be achieved exactly once, by a fine-tuning of the free parameter 
\begin{equation}\label{unique}
c\to c_P\,,
\end{equation} 
see \eq{cP}, \eq{asymp}. This is equivalent to the fine-tuning of the exactly marginal sextic coupling $\tau\to \tau_P$, \eq{BMB}. In view of the discussion in Sec.~\ref{SBSI} we conclude that the role of the  free dimensionless parameter $c$ which determines the exactly marginal sextic coupling $\tau$,  is replaced by a free dimensionful parameter determining the mass term $m^2$ in the symmetric phase. The phenomenon whereby the role of a dimensionless parameter is taken over by a dimensionful one, due to fluctuations, is known as ``dimensional transmutation'' \cite{Coleman:1973jx}. We conclude that dimensional transmutation is operative exactly at the critical coupling strength \eq{unique}, accompanied by the appearance of a dilaton \cite{Bardeen:1983rv,Bardeen:1983st}.

A divergence of the dimensionless mass term at vanishing field \eq{uprime0} cannot arise through a fixed point potential which is an analytic function of the field. For dimensional transmutation to become operative through \eq{uprime0}, the theory must develop some non-analyticities at small field. For the choice $c=c_P$ \eq{cP}, and using the explicit solution, we find that the mass function $u'_*(\rho)$ at the BMB fixed point potential displays a root-type non-analyticity $u'_*\propto \, 1/{\sqrt{\rho}}$ for small field, leading to
\begin{eqnarray}\label{div}
u_{\rm BMB}' \, =\, \frac{1}{\sqrt{5\rho}}-\frac{5}{7}+\frac{10}{147} \sqrt{5\rho}
+\frac{800}{11319}\,\rho+{\cal O}(\rho^{3/2})\,.
\end{eqnarray}
For all other choices $c\neq c_P$ of the coupling parameter the potential remains smooth and analytical for small positive $\rho$. The non-analytic behaviour \eq{div} does not entail any divergence for the potential itself.  In fact, the potential remains bounded even at vanishing field, 
\begin{eqnarray}\label{BMBsq}
u_* \, \propto \, \sqrt{\rho}\,,
\end{eqnarray}
showing that the theory remains well-defined even though a non-analyticity has arisen through quantum fluctuations at strong coupling.

However, we stress that our result is in marked contrast with
the shape of the effective potential given in \cite{David:1984we} where a non-analyticity of the form
\begin{eqnarray}\label{BMBdiv}
u_*' \, \propto \, \frac{1}{\rho}
\end{eqnarray}
has been observed. In contradistinction to \eq{BMBsq}, the resulting effective potential is no longer bounded from below \cite{David:1984we},
\begin{eqnarray}\label{BMBln}
u_* \, \propto \, \ln{\rho}\,.
\end{eqnarray}
It has also been argued that the logarithmic divergence in \eq{BMBln} is weak enough to allow 
for a meaningful ground state \cite{David:1984we}. We note that effective potentials are not observable and their precise shapes can depend on unphysical parameters such as the regularisation scheme.
In fact, the origin for the difference 
between \eq{BMBsq} and \eq{BMBln} is that different RG schemes have been adopted for the regularisation of the theory, a smooth optimised scheme for \eq{BMBsq}, and the sharp cutoff scheme for \eq{BMBln}. In this light, the important question arises whether the observed phenomenon is physical, or rather an artefact of the underlying regularisation.
To answer this question,
we must study the RG scheme dependence of the BMB potential in more detail, to which we turn next.

\subsection{Universality of the BMB phenomenon}
We now clarify to which extend the RG scheme is responsible for the non-analytical  behaviour at the BMB fixed point, and how the characteristics of the BMB critical potential depend on it.  To that end, we introduce the RG flow for a more general set of cutoffs, interpolating between the optimised cutoff used above, the sharp cutoff as used in the original BMB study. Following  \cite{Litim:2005us}, and using a normalisation of the fields and the potential such that the VEV remains at $\kappa=1$, we define a family of RG schemes via a family of  RG flows given by
\begin{equation}\label{gammaflow}
\partial_t u'=-2u'+\rho u''-\frac{u''}{(1+u')^\gamma}\,.
\end{equation}
A few comments are in order. The parameter $\gamma=\gamma(R_k)$ has become a placeholder for the RG scheme. For specific values, the corresponding regulator function $R_k$ is known explicitly. The  value $\gamma_{\rm opt}=2$ corresponds to the optimised cutoff $R_{\rm opt}$ given in \eq{opt} \cite{Litim:2000ci,Litim:2001up}. The sharp cutoff $R_s=\lim_{a\to\infty} a \cdot\theta(k^2-q^2)$ corresponds to  $\gamma_s=1$, and $\gamma_q=3/2$ to an algebraic cutoff quartic in momenta $R_q\propto q^4/k^2$. With a modified normalisation of the fields, the limit $\gamma\to 0$ can formally also be taken, leading to a logarithmic term. It corresponds to a plain mass term cutoff where $R_k\propto k^2$, and a flow of the Callan-Symanzik-type. 
 More generally, for functional flows derived from \eq{FRG}, the parameter $\gamma$ can in principle vary between $\gamma\in [0,2]$. Low values $\gamma\in [0,1)$ lead to less stable flows and require additional care \cite{Litim:2002hj}, and we will restrict ourselves to the range $\gamma\in [1,2]$. In a background field formulation of the Wetterich flow,  the parameter $\gamma$ may take formally any real positive value \cite{Litim:2001up,Litim:2001hk,Litim:2001dt}. This limit reduces to the so-called proper-time flow in the approximation adopted here \cite{Bohr:2000gp,Bonanno:2000yp}. The interpretation of the RG flow is modified due to an implicit re-organisation of diagrams due to the presence of background fields.

\begin{figure}[t]
\begin{center}
\includegraphics[scale=0.75]{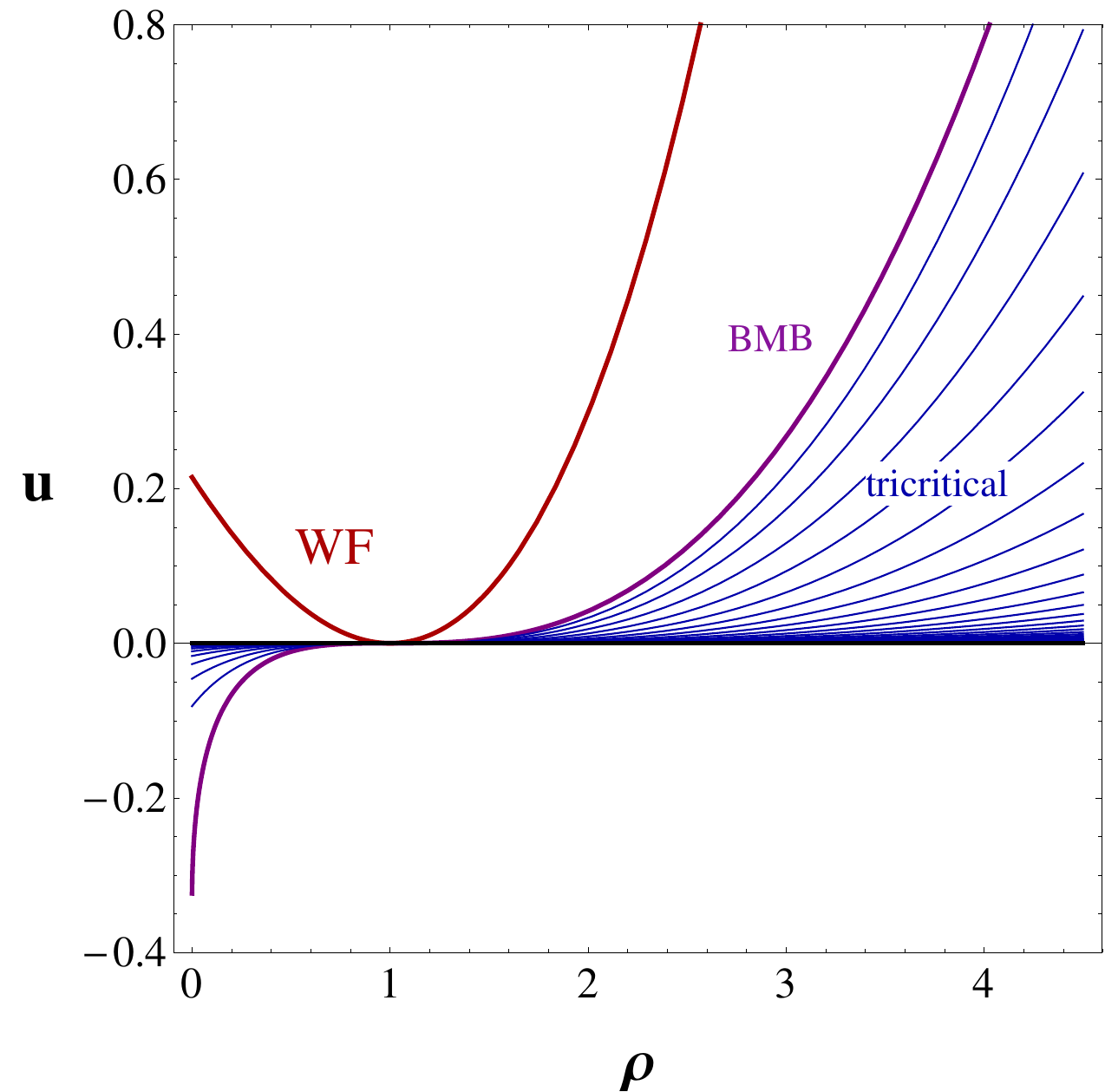}
\end{center}
\caption{The shape of the dimensionless fixed point potential $u$ as a function of the field $\rho$ for different parameters $c$. The Wilson-Fisher fixed potential (WF, red line) has a global minimum at $\rho=\kappa$. In contrast, the tricritical fixed points $(c>c_P,$ blue lines) and the BMB solution $(c=c_P,$ pink line) display a saddle point at  $\rho=\kappa$. The BMB solution becomes non-analytical at vanishing field.
\label{potplot}}
\end{figure}

The flow \eq{gammaflow} can be integrated analytically, providing us with an explicit solution of the form \eq{+} and the function $F$, now expressed in terms of the hypergeometric series
\begin{equation}\label{Fgamma}
F(u')=\frac{1}{\sqrt{u'}}\ {}_2F_1(-\s012,\gamma,\s012,-u')\,.
\end{equation}
Similar results are found for the branch \eq{-} with negative $u'$. 
For $\gamma=2$ these fall back onto the results \eq{F+} and \eq{F-}. For large $u'$, a convenient representation is given by
\begin{eqnarray}\label{Fgamma2}
F(u')&=&
\sqrt{u'}\cdot \Gamma(\s012)\frac{\Gamma(\s012+\gamma)}{\Gamma(\gamma)}
 +\frac{(u')^{-\gamma}}{2\gamma+1}\cdot
{}_2F_1\left(\gamma,\s012+\gamma,\s032+\gamma,-\s0{1}{u'}\right)\,,
\end{eqnarray}
showing that the large-$u'$ asymptotics is proportional to $\sqrt{u'}$. For each of these solutions, we then determine the specific parameter $c$ corresponding to the BMB fixed point. For large $u'$, it can be read off as the coefficient of the term linear in $\sqrt{u'}$ from \eq{Fgamma2}, hence
\begin{equation}\label{c*}
c_P=\frac{\Gamma(\s012)\,\Gamma(\s012+\gamma)}{\Gamma(\gamma)}\,.
\end{equation}
For generic RG scheme, we have thus established that there exists exactly one critical value \eq{c*} for the coupling parameter leading to the BMB behaviour. We recall that the BMB solution is characterised by the divergence of $u'\to\infty$ in the limit $\rho\to 0$. At the BMB point, and in the vicinity of vanishing field, we  find that $u'$ diverges with a scheme-dependent inverse power of the field
 \begin{equation}\label{u'}
 u'(\rho)=[(1+2\gamma)\rho]^{-1/\gamma}+{\rm subleading}\,.
 \end{equation}
 We notice that the divergence in $u'$ for small fields at the critical coupling \eq{c*} is a stable feature of the theory, irrespective of the choice for the RG scheme, although the details of the non-analyticity depend on it. 
 
 \subsection{Boundedness of the effective potential}
 We are now in a position to address the boundedness of the effective potential. The expression \eq{u'} can be integrated. Omitting constant terms, the explicit expression for the non-analytical behaviour of the BMB effective potential for small fields reads
 \begin{equation}\label{non}
 u(\rho)=\gamma\,(1+2\gamma)^{-1/\gamma}\,\frac{\rho^{1-1/\gamma}}{\gamma-1}
 +{\rm subleading}
\,. \end{equation}
We notice that the details of the non-analyticity  $u\sim\rho^{1-1/\gamma}$ depend on the  momentum cutoff $\gamma(R_k)$.
 For the optimised cutoff $(\gamma=2)$, and apart from an overall normalisation constant for the vacuum energy, the non-analyticity \eq{non} is of the square root-type, $u=2\sqrt{\rho}/\sqrt{5}$ \eq{div} modulo subleading corrections in the field. In the sharp cutoff limit $(\gamma\to 1)$, we reobtain a logarithmic divergence of the form $u=\s013 \ln\rho$, \eq{BMBln}, again up to subleading terms and dropping constant ones. The important difference is that while the potential at vanishing field remains finite for the optimised cutoff, it becomes unbounded from below for the sharp cutoff. 
 Next we introduce the depth of the potential defined as the difference between its value at the saddle point $\rho=\kappa$ and its value at vanishing field
\begin{equation}\label{delta}
\Delta=
{u(\kappa)-u(0)}\,.
\end{equation}
Using the relations \eq{1-}, \eq{H} together with $c=c_P$ for the inner part of the potential, and expressing $\rho$ and $u''$ as functions of $x\equiv u'$, the potential depth \eq{delta} can be rewritten as
 an integral
 \begin{equation}\label{diff}
 \Delta=\frac 12 \int_0^\infty dx \sqrt x \left[c_P-F(x)-2 x F'(x)\right]\,.
 \end{equation}
Inserting the explicit expressions for $F$ and $c_P$ as in \eq{Fgamma}, \eq{c*}, the potential difference \eq{diff} is obtained analytically and reads
 \begin{equation}\label{Delta}
 \Delta=\frac{1}{3(\gamma-1)}
 \end{equation}
as a function of the scheme parameter. For large $\gamma\to\infty$, the potential difference becomes parametrically small. In turn, for $\gamma\to 1$ the potential difference diverges. 

We can now make the following observations. Firstly, for generic regularisation, the BMB fixed point potential comes out bounded rather than unbounded. Secondly, the potential difference is finite (in units of the RG scale) for generic RG scheme, except for the extreme limit of a sharp cutoff. The previously detected unboundedness must therefore be understood as an artefact of the sharp cutoff regularisation, rather than an artefact of the physical theory. In the physical  limit \eq{k0} the metastable vacuum state with  $\bar \kappa=\kappa\cdot k\to 0$ coincides with the true vacuum at vanishing field, and the potential difference 
vanishes, $k^3\cdot \Delta \to 0$ for $k\to 0$. 
In Fig.~\ref{potplot}, the (dimensionless) effective potential at the fixed point is shown for the optimised cutoff, and as a function of the sextic coupling. Evidently, all dimensionless  tricritical potentials display a metastable extremum at $\kappa=1$ and a global minimum in the symmetric phase where $\kappa=0$. In turn, the global minimum is located at $\kappa=1$ for the Wilson-Fisher fixed point.

 \subsection{$1/N$ corrections}\label{1/N}

Finally, we comment on corrections beyond the leading order in $1/N$. 
It is well-known that non-analyticities such as those in \eq{div} cannot arise through  Feynman diagrams at any finite order in perturbation theory. One may then wonder whether the BMB phenomenon is a genuine strong coupling phenomenon or a consequence of the infinite $N$ limit. In fact, it has already been argued that the BMB phenomenon is no longer operative as soon as $N$ takes finite values \cite{David:1985zz,Omid:2016jve,Heilmann:2012yf}. More recently,  it has also been established that scaling potentials at finite $N$ do not converge pointwise to those at infinite $N$  at the Wilson-Fisher fixed point, for all fields \cite{Juttner:2017cpr}. On a technical level, once $N$ is finite, radial mode fluctuations and derivative interactions are switched on and start to compete with the Goldstone mode fluctuations.

Here, we  investigate the fate of the BMB phenomenon by exploring
the competition between radial and Goldstone mode fluctuations to leading order in the derivative expansion 
\cite{Litim:2016hlb} (see \cite{Heilmann:2012yf} for a similar study in supersymmetric models). We have seen in \eq{uprime0}  that the very existence of the BMB phenomenon requires the divergence of the mass function $u'_*(\rho)$ at vanishing field. At infinite $N$, this is dictated by the Goldstone mode fluctuations.
At the BMB fixed point, they lead to a  mass function of the form \eq{div}
for small fields. It is straightforward to confirm that \eq{div} solves the fixed point condition $\partial_t u'=0$, \eq{duprime}, for each and every order as a power series in $\sqrt{\rho}$. On the level of the flow, the leading divergence $\propto 1/\sqrt{\rho}$ arises from the canonical scaling and is exactly balanced by fluctuations. Let us now consider  $1/N$ corrections whereby the RHS of \eq{duprime} becomes
\begin{equation}\label{duBMB}
\partial_t u'= -2u'+\rho u'' -\frac{u''}{(1+u')^2}-\frac{1}{N-1}\frac{3u''+2\rho u'''}{(1+u'+2\rho u'')^2}
\,.
\end{equation}
The last term accounts for the fluctuations of the radial mode. 
We observe that the  radial numerator $(3u''+2\rho u''')$ vanishes identically for the leading divergence $u'\propto {\rho}^{-1/2}$ of the BMB solution. On the other hand, the radial denominator $(1+u'+2\rho u'')^2\propto\rho^{-3}$ diverges more strongly than the Goldstone denominator $(1+u')^2\propto \rho^{-1}$. In combination, this implies that the leading and subleading $1/N$ behaviour is the same.
We therefore may search for a BMB type solution which is parametrically close to  \eq{div} in $1/N$, and which starts out as  $\sim \rho^{-1/2}$. We find
\begin{eqnarray}\label{divN}
u_*' \, &=&
+ \left[\frac{1}{\sqrt{5\rho}}-\frac{5}{7}+\frac{10}{147} \sqrt{5\rho}
+\frac{800}{11319}\,\rho+\cdots\right]
\nonumber\\ &&
- \left[\frac56\frac{1}{\sqrt{5\rho}}
+\frac{100}{693}
-\frac{249077}{2083158}\sqrt{5\rho}
+\frac{317711300}{2058507297}
\,\rho+\cdots\right]\, \frac1N
\\ &&
 -\left[
\frac{1662097}{595188}\frac{1}{\sqrt{5\rho}}
-\frac{66240095}{31696434}
+\frac{23209996241}{37053131346}
\sqrt{5\rho}
+\frac{26372394267098720}{23585077690921593}
\,\rho+\cdots\right]\, \frac1{N^2} 
 \nonumber \\ && \nonumber
+\, {\cal O} (\rho^{3/2})\,+\, {\cal O}\left({1/N^{3}}\right)
\end{eqnarray}
which solves the fixed point condition $\partial_t u'=0$ with \eq{duBMB} up to corrections of order ${\cal O}\left({1/N^{3}}\right)$ and ${\cal O}\left(\rho^{3/2}\right)$.  At infinite $N$, the solution \eq{divN} extends over all fields. It remains to be clarified whether the same holds true at finite $N$. Due to the competition between radial and Goldstone mode fluctuations this is not the case in the supersymmetric version of the model \cite{Heilmann:2012yf}. We  therefore should expect that the series \eq{divN} terminates at a singularity for finite fields. This is left for future work. In a similar vein, it has been noted that the Wilson-Fisher fixed point at finite $N$ does not converge pointwise to the infinite $N$ result for all fields \cite{Juttner:2017cpr}. It would seem interesting to clarify the status of $1/N$ corrections for the tricritical fixed points along the same lines.

\subsection{Discussion}
Using functional renormalisation, we have established the  existence of a unique non-perturbative fixed point corresponding to the Bardeen-Moshe-Bander phenomenon  irrespective of the underlying RG scheme. The fixed point  invariably arises through a non-analyticity of the scalar potential at small fields, thereby ensuring \eq{uprime0}. The precise form of the non-analyticity at vanishing field is non-universal in that it depends on technical parameters such as the Wilsonian momentum cutoff. For generic regularisation, the resulting effective potential is genuinely bounded from below except for singular  choices of the RG scheme such as the notorious sharp momentum cutoff.  We also stress that our approximation has become exact owing to the infinite $N$ limit in the absence of derivatve interactions.  We can therefore safely conclude that the Bardeen-Moshe-Bander phenomenon is a feature of the physical theory at infinite $N$, and not an artefact of the underlying regularisation.

\section{\bf Summary}\label{C}

Exactly solvable models are hard to come by, but once available, they offer important analytical insights into the inner working of quantum field theory. We have investigated interacting fixed points of the exactly solvable  $O(N)$-symmetric scalar $(\phi^2)^3_{d=3}$ model at infinite $N$, using functional renormalisation, Figs.~\ref{ccode}~,\ref{fixed}.
A central role is played by the sextic scalar selfcoupling, which becomes exactly marginal for vanishing quartic interactions. For weak sextic coupling, the theory displays a line of interacting UV fixed points starting out of the Gaussian fixed point,
Fig.~\ref{branchweak}. With increasing sextic coupling, the potential ceases to exist beyond a critical value. Non-analyticities develop in the small field region leading to a Landau-type singularity in the quartic at strong coupling, Fig.~\ref{branch}.  Interestingly though, even at strong coupling the theory displays classical scaling at the ``would-be'' fixed point despite of interactions. 
Another novelty is the phase diagram Fig.~\ref{pBMB} showing  how the asymptotically safe UV fixed points are connected with the Wilson-Fisher fixed point and the symmetric and symmetry broken phases at low energies.

Non-classical scaling arises at the Wilson-Fisher fixed point, an isolated scaling solution which cannot be reached from the tricritical line by continuous tuning of the sextic coupling. 
Also, at a critical value for the sextic, the theory displays the spontaneous breaking of scale invariance through the generation of mass. Its fingerprint is a non-analyticity of the effective potential. Although the details for the latter depend on the regularistion, the phenomenon arises universally. We have also established that the BMB potential remains bounded irrespective of the regularisation scheme, Fig.~\ref{potplot}, with the exception of the notorious sharp cutoff where we confirm logarithmic unboundedness.  First steps have been made to extend the BMB solution beyond infinite $N$, though more work is required to relate with the results of  \cite{David:1985zz,Omid:2016jve,Heilmann:2012yf}.

On a technical level, we have observed that  classical and non-classical scaling exponents arise correctly from both the local and the global RG flows. The critical endpoint 
and the onset of strong coupling
is also well-captured by the global RG flows. Local RG flows, on the other hand,  
did not offer good indications for the endpoint, reminiscent of large-$N$ perturbation theory.
We conclude that in order to uncover the onset of strong coupling, non-perturbative methods beyond polynomial expansions or conventional perturbation theory should be adopted.  
\\[3ex]

\newpage
\centerline{\bf Acknowledgements}
${}$\\[-2ex]
We thank Kevin Falls, Yannick Meurice, and Rob Pisarski for useful discussions. DL also thanks the Aspen Center for Physics for hospitality and its stimulating environment which initiated this project.  MP thanks the University of Sussex for hospitality during an extended research visit. This work is supported by the Science and Technology Facilities
Council (STFC) under grant number ST/L000504/1, and by the Hungarian Research Fund (OTKA) under contracts No.~K104292.

\bibliography{biblio2_modif,biblio2_modif2}
\bibliographystyle{JHEP}

\end{document}